\documentclass[preprint]{ptephy_v1}%%%%%% to generate preprint number
\usepackage{braket}
\usepackage{comment}
\usepackage{xcolor}

\preprintnumber{XXXX-XXXX} %%% %%% Insert preprint number here

\def\Journal#1#2#3#4{{#1} {\bf #2}, #3 (#4)}
\def\ARNPS{Annu. Rev. Nucl. Part. Sci.} 
\def\AandA{Astron. Astrophys.} 

\def\APJ{Astrophys. J.}

\def\CMP{Commn. Math. Phys.}

\def\CQG{Class. Quant. Grav.}

\def\EPJP{Eur. Phys. J. Plus}

\def\IJMPA{Int. J. Mod. Phys. A}
\def\IJMPD{Int. J. Mod. Phys. D}

\def\JCAP{J. Cosmol. Astropart. Phys.}
\def\JHEP{J. High Energy Phys.}
\def\JETP{J. Exp. Theor. Phys}

\def\MNRAS{Mon. Not. R. Astron. Soc.}

\def\NATULONDON{Nature (London)}

\def\PLB{{Phys. Lett.} B}

\def\PRD{Phys. Rev. D}

\def\PTP{Prog. Theor. Phys.}
\def\PTEP{Prog. Theor. Exp. Phys.}

\def\RPP{Rep. Prog. Phys.}

\begin{document}

\title{Primordial black holes and dark matter mass spectrum}

%%%% To generate auto affiliation numbers please use \author{}\affil{} command

\author{Teruyuki Kitabayashi}
\affil{Department of Physics, Tokai University,
4-1-1 Kitakaname, Hiratsuka, Kanagawa 259-1292, Japan\email{teruyuki@tokai-u.jp}}

%%% To include the collaborator name... Please use the command "\collaborator"
%%% For example: \collaborator{ATLAS Collaboration}

\begin{abstract}
Because primordial black holes (PBHs) evaporate into all particle species in nature, PBHs may emit several dark matter (DM) particle species with specific mass spectra. We assume that PBHs are the only source of DMs, and DMs only interact with the standard model particles gravitationally. We show a relation between the number of DM particle species $N_{\rm DM}$ and initial PBH density $\beta$ and mass $M_{\rm BH}^{\rm in}$. $\beta$-$M_{\rm BH}^{\rm in}$ curves for different $N_{\rm DM}$ tend to overlap with each other for heavy initial PBHs. We also show that the allowed region of DM masses for multiple DM.
\end{abstract}

\subjectindex{}

\maketitle

%%----------------------------------------------------------------------------------
\section{Introduction\label{section:introduction}}
%%----------------------------------------------------------------------------------
A primordial black hole (PBH) is a kind of black hole that could have been formed in the early Universe. PBHs are receiving a lot of attention even though there is no clear evidence of their existence. For a recent review, see \cite{Carr2010PRD,Carr2020ARNPS,Carr2021RPP}.
 
One of the reasons PBHs are attracting attention is that their existence could lead to a solution to the dark matter (DM) problem in cosmology. DM is an unknown gravitational source in the Universe. DM may be made of single or multiple particle species. Because PBHs emit particles via the Hawking radiation induced by gravity \cite{Hawking1975CMP},  PBHs evaporate into all particle species in nature. The resultant Hawking radiation of PBHs is a possible explanation for DM production \cite{Bell1999PRD,Green1999PRD,Khlopov2006CQG,Baumann2007arXiv,Dai2009JCAP,Fujita2014PRD,Allahverdi2018PRD,Lennon2018JCAP,Morrison2019JCAP,Hooper2019JHEP,Masina2020EPJP,Baldes2020JCAP,Bernal2021JCAP,Gondolo2020PRD,Bernal2020arXiv2,Auffinger2020arXiv,Datta2021arXiv,Chaudhuri202011arXiv,Sandick2021PRD,Cheek2022PRD,Baker2022arXiv,Kitabayashi2021IJMPA,Kitabayashi2022PTEP,Kitabayashi2022arXiv2208.09566}. 

In the literature, it is typically assumed that a single DM particle species is produced via PBH evaporation. However, because PBH evaporation would produce any particle species in principle, several DM species with a specific mass spectrum may be emitted from PBHs \cite{Cheek2022PRD,Baker2022arXiv}. 

The phenomenology of PBHs with multiple DMs has already been studied in Refs. \cite{Bell1999PRD, Cheek2022PRD,Baker2022arXiv}. Some connections between the evaporation of PBHs and mirror matters as DMs have been explored by Bell and Volkas in Ref. \cite{Bell1999PRD}. They found a connection between the effective number of neutrinos and PBHs with multiple DM sectors; however, because they included the effects of DM sectors in their analysis via the total energy density of the mirror matter sector, there is no discussion about DM mass spectrum in Ref. \cite{Bell1999PRD}. In Ref. \cite{Cheek2022PRD}, Cheek et al., demonstrated useful analytical formulae for estimating DM relic abundance from BPH evaporation that included the effect of the greybody factor (they also showed the result from precise numerical calculations). In addition, they assumed the existence of additional and unstable degrees of freedom emitted by the evaporation of PBHs, which later decay into DMs. Because they estimated the final relic abundance of DM with only one additional heavy particle, there was no discussion on the DM mass spectrum in detail in Ref. \cite{Cheek2022PRD}. Baker and Thamm proposed a new method for probing the particle spectrum of nature with evaporating PBH remnants in today's Universe in Ref. \cite{Baker2022arXiv}. They assumed that the dark sector consists of N copies of standard model particles and that all particles in the dark sector have common mass. For a PBH to survive in today's Universe, the initial PBH mass should be approximately $10^{15}$ g. Thus, only a case in which all DM species degenerate with the same mass and PBHs have very large initial mass of approximately $10^{15}$ g was mainly discussed in Ref. \cite{Baker2022arXiv}. The discussions in these three interesting papers motivated us to start our study.

In this paper, we study the effect of the DM mass spectrum on the PBH and vice versa. The following four specific mass spectra of the DM sector are considered:
\begin{itemize}
 \item Single DM with a mass $m_{\rm DM}$.
 \item Multiple DMs with a degenerated mass spectrum
 \begin{equation}
 \underbrace{m, m,  m, \cdots, m}_{N_{\rm DM}}. \nonumber 
 \end{equation}
 \item Multiple DMs with an arithmetic sequence-like mass spectrum
 \begin{equation}
  \underbrace{m_0, m_0+\Delta m,  m_0+2\Delta m, \cdots, M}_{N_{\rm DM}}, \nonumber 
 \end{equation}
 where $m_0$ and $M$ denote the minimum and maximum mass in the mass spectrum, respectively. $\Delta m$ denotes the tolerance.
 \item Multiple DMs with a geometric sequence-like mass spectrum
 \begin{equation}
  \underbrace{m_0, m_0 r,  m_0r^2, \cdots, M}_{N_{\rm DM}}, \nonumber 
 \end{equation}
\end{itemize}
where $r$ denotes the geometric ratio.

This paper is organized as follows. In Sec. \ref{sec:PBH}, we present a review of PBHs.  Sec. \ref{sec:PBH} is mainly based on Refs. 
\cite{Fujita2014PRD,Morrison2019JCAP,Masina2020EPJP,Gondolo2020PRD,Bernal2021JCAP,Cheek2022PRD}. In Sec. \ref{section:PBHandDM}, we show the effect of DM mass spectra on the initial PBH density. In Sec. \ref{sec:four_cases}, we estimate the allowed region of DM mass. Section \ref{sec:summary} summarizes the study.

We use the natural unit ($c=\hbar=k_{\rm B}=1$) in this paper.

%%----------------------------------------------------------------------------------
\section{Primordial black holes\label{sec:PBH}}
%%----------------------------------------------------------------------------------

%%----------------------------------------------------------------------------------
\subsection{Early Universe}
%%----------------------------------------------------------------------------------
We discuss some fundamental knowledge of the early Universe. Our discussion is based on the cosmological principle, that the Universe is homogeneous and isotropic, and we neglect the curvature and cosmological constant in the early Universe. The evolution of the homogeneous and isotropic Universe is described using the Friedmann equation 
\begin{eqnarray}
\left(\frac{\dot{a}(t)}{a(t)} \right)^2 = H(t)^2 = \frac{8\pi G}{3} \rho(t) ,
\label{Eq:Friedmann}
\end{eqnarray}
where $a(t)$, $H(t)$, $G$, and $\rho(t)$ denote the scale factor, Hubble parameter, gravitational constant, energy density of the Universe, respectively.

In the early Universe, the main ingredients were relativistic particles (these particles are called radiation). The energy density of radiation is given by
\begin{eqnarray}
\rho_{\rm rad}(T) = \frac{\pi^2}{30} g_*(T)T^4,
\label{Eq:rho_rad}
\end{eqnarray}
where
\begin{eqnarray}
g_*(T) = \sum_{i={\rm bosons}}g_i \left(\frac{T_i}{T}\right)^4+ \frac{7}{8} \sum_{i={\rm fermions}}g_i \left(\frac{T_i}{T}\right)^4,
\end{eqnarray}
denotes the relativistic effective degrees of freedom for the radiation energy density, and $T$ denotes the temperature of the Universe. In the radiation-dominated era, we have $\rho_{\rm rad} \propto a^{-4}$,  $a \propto t^{1/2}$, and the Hubble parameter is obtained as 
\begin{eqnarray}
H(T) = \sqrt{\frac{4\pi^3G}{45}}g_*(T)^{1/2}T^2 = \frac{1}{2t},
\end{eqnarray}
using Eqs. (\ref{Eq:Friedmann}), (\ref{Eq:rho_rad}). The temperature-time relation is
\begin{eqnarray}
T(t) = \left( \frac{45}{16\pi^3G} \right)^{1/4}\frac{1}{g_*(T)^{1/4} t^{1/2}}.
\label{Eq:temp_time_RD}
\end{eqnarray}

The entropy density is given by
\begin{eqnarray}
s(T) = \frac{2\pi^2}{45} g_{*s}(T)T^3,
\label{Eq:s}
\end{eqnarray}
where
\begin{eqnarray}
g_{*s}(T) = \sum_{i={\rm bosons}}g_i \left(\frac{T_i}{T}\right)^3+ \frac{7}{8} \sum_{i={\rm fermions}}g_i \left(\frac{T_i}{T}\right)^3,
\end{eqnarray}
denotes the  relativistic effective degrees of freedom for the entropy density.

%%----------------------------------------------------------------------------------
\subsection{PBH formation}
%%----------------------------------------------------------------------------------
PBHs are produced in the early Universe through several mechanisms, such as the collapse of large density perturbations generated from inflation \cite{Garcia-Bellido1996PRD,Kawasaki1998PRD,Yokoyama1998PRD,Kawasaki2006PRD,Kawaguchi2008MNRAS,Kohri2008JCAP,Drees2011JCAP,Lin2013PLB,Linde2013PRD}, a sudden reduction in the pressure \cite{Khlopov1980PLB,Jedamzik1997PRD}, bubble collisions \cite{Crawford1982NATULONDON,Hawking1982PRD,Kodama1982PTP,La1989PLB,Moss1994PRD}, a curvaton \cite{Yokoyama1997AandA,Kawasaki2013PRD,Kohri2013PRD,Bugaev2013IJMPD}, and collapse of cosmic string \cite{Hogan1984PLB}.

We assume that PBHs are produced in the early Universe by large density perturbations generated from inflation \cite{Garcia-Bellido1996PRD,Kawasaki1998PRD,Yokoyama1998PRD,Kawasaki2006PRD,Kawaguchi2008MNRAS,Kohri2008JCAP,Drees2011JCAP,Lin2013PLB,Linde2013PRD}, PBH mass is proportional to horizon mass, and PBHs have the same masses at their formation times. In addition, we assume that PBHs form during the radiation-dominated era with a monochromatic mass function, and PBHs do not have angular momentum and electric charge (we study Schwarzschild PBHs in this paper).

The initial mass of a PBH formed in a radiation-dominated era is evaluated as 
\begin{eqnarray}
M_{\rm BH}^{\rm in} = \frac{4\pi}{3}\gamma \rho_{\rm rad} (T_{\rm in}) H(T_{\rm in})^{-3},
\label{Eq:PBHformationMin}
\end{eqnarray}
where $T_{\rm in}$ denotes the temperature of the Universe at the PBH formation time, and $\gamma$ denotes a numerical factor that depends on the details of the gravitational collapse. According to Carr's formula, $\gamma \sim 0.2$ \cite{Carr1975APJ}.  The temperature of the Universe at the PBH formation time is obtained as 
\begin{eqnarray}
T_{\rm in} = \left(\frac{45}{16\pi^3}\right)^{1/4} \frac{\gamma^{1/2}}{g_*(T_{\rm in})^{1/4}} \left( \frac{M_{\rm Pl}^3}{M_{\rm BH}^{\rm in}} \right)^{1/2},
\label{Eq:Tin}
\end{eqnarray}
using Eqs. (\ref{Eq:Friedmann}), (\ref{Eq:rho_rad}), and (\ref{Eq:PBHformationMin}), where $M_{\rm Pl}=\sqrt{1/G}$ is the Planck mass.

We introduce the dimensionless parameter
\begin{eqnarray}
\beta = \frac{\rho_{\rm BH}^{\rm in}}{\rho_{\rm rad} (T_{\rm in})} = \frac{M_{\rm BH}^{\rm in} n_{\rm BH}(T_{\rm in})}{\rho_{\rm rad} (T_{\rm in})},
\label{Eq:beta}
\end{eqnarray}
to represent the initial energy density of PBHs at the time of their formation.

%%----------------------------------------------------------------------------------
\subsection{PBH evaporation}
%%----------------------------------------------------------------------------------
A black hole loses its mass by producing particles with masses below the Hawking temperature (horizon temperature of the black hole)
\begin{eqnarray}
T_{\rm BH} = \frac{1}{8\pi G M_{\rm BH}},
\end{eqnarray}
via Hawking radiation \cite{Hawking1975CMP}. The emission rate of particle species $i$ is expressed as 
\begin{eqnarray}
\frac{d^2 \mathcal{N}_i}{dtdE} = \frac{g_i}{2\pi} \frac{\Gamma_i(E, M_{\rm BH})}{e^{E/T_{\rm BH}}-(-1)^{2s_i}}, 
\end{eqnarray}
where $\mathcal{N}_i$ denotes the number of particle species $i$, $E$ denotes the energy of the particle, $M_{\rm BH}$ is the black hole mass, $g_i$, and $s_i$ denote the number of degrees of freedom and spin of particle $i$, respectively, and $\Gamma_i$ denotes the greybody factor. According to the parametrization of the greybody factor proposed by Cheek et.al, \cite{Cheek2022PRD}, we obtain
\begin{eqnarray}
\frac{d^2 \mathcal{N}_i}{dtdE} = \frac{g_i}{2\pi} \frac{27 G^2 M_{\rm BH}^2 \psi_{s_i}(E) (E^2-m_i^2)}{e^{E/T_{\rm BH}}-(-1)^{2s_i}}, 
\end{eqnarray}
where $m_i$ denotes the mass of particle $i$ and $\psi_{s_i}$ is the absorption cross-section normalized to the geometric optics limit.

The time evolution of the black hole mass due to Hawking radiation is given by 
\begin{eqnarray}
\frac{dM_{\rm BH}}{dt} = -\sum_i\int_0^\infty \frac{d^2\mathcal{N}_i}{dtdE}EdE 
 = -\sum_i g_i\epsilon_i(z_i)\frac{M_{\rm Pl}^4}{M_{\rm BH}^2},
\end{eqnarray}
where
\begin{eqnarray}
\epsilon_i(z)&=&\frac{27}{8192\pi^5}\int_{z}^\infty \frac{\psi_{s_i}(x)(x^2-z^2)}{e^x-(-1)^{2s_i}}xdx  \\
&\simeq&a_\epsilon\{ 1-\left[1+\exp(-b_\epsilon\log_{10}z + c_\epsilon) \right]^{-d_\epsilon} \}, \nonumber
\end{eqnarray}
$z_i=m_i/T_{\rm BH}$ and \cite{Cheek2022PRD}
\begin{eqnarray}
a_\epsilon = 
\begin{cases}
7.61 \times 10^{-5} & ({\rm scalar}) \\
4.12 \times 10^{-5} & ({\rm fermion}) \\
1.68 \times 10^{-5} & ({\rm vector}) \\
1.93 \times 10^{-6} & ({\rm graviton}) \\
\end{cases},
\quad
b_\epsilon = 
\begin{cases}
7.79884 & ({\rm scalar}) \\
13.0496 & ({\rm fermion}) \\
14.0361 & ({\rm vector}) \\
21.5094 & ({\rm graviton}) \\
\end{cases},
\end{eqnarray}
\begin{eqnarray}
c_\epsilon = 
\begin{cases}
3.80742 & ({\rm scalar}) \\
9.91178 & ({\rm fermion}) \\
10.7138 & ({\rm vector}) \\
20.5135 & ({\rm graviton}) \\
\end{cases},
\quad
d_\epsilon = 
\begin{cases}
0.4885 & ({\rm scalar}) \\
0.3292 & ({\rm fermion}) \\
0.3072 & ({\rm vector}) \\
0.1734 & ({\rm graviton}) \\
\end{cases}.
\end{eqnarray}

The lifetime of the black hole can be evaluated as 
\begin{eqnarray}
\tau = t_{\rm ev} - t_{\rm in} 
= \int_{t_{\rm in}}^{t_{\rm ev}} dt 
= \frac{1}{M_{\rm Pl}^4} \int_0^{M_{\rm BH}^{\rm in}}   \frac{M_{\rm BH}^2}{\sum_i g_i\epsilon_i(z_i)}   dM_{\rm BH}.
\end{eqnarray}

We assume that all PBHs evaporated completely during the radiation-dominant era. In this case, the condition
\begin{eqnarray}
\beta < \beta_c = \frac{T_{\rm ev}}{T_{\rm in}},
\end{eqnarray}
should be satisfied \cite{Fujita2014PRD, Morrison2019JCAP, Masina2020EPJP,Bernal2021JCAP,Gondolo2020PRD,Cheek2022PRD}, where 
\begin{eqnarray}
T_{\rm ev} \simeq T(\tau) = \left( \frac{45}{16\pi^3} \right)^{1/4}\frac{1}{g_*(T_{\rm ev})^{1/4}}  \left(\frac{M_{\rm Pl}}{\tau}\right)^{1/2}, 
\label{Eq:Tev}
\end{eqnarray}
denotes the temperature of the Universe at the time of PBH evaporation.

The total number of particle species $i$ emitted from a single black hole is
\begin{eqnarray}
\mathcal{N}_i &=& \int_0^\tau dt \int_0^\infty \frac{d^2 \mathcal{N}_i}{dtdE} dE \\
&=& \frac{27g_i}{1024\pi^4(z_i^{\rm in})^2}\frac{\left(M_{\rm BH}^{\rm in}\right)^2}{M_{\rm Pl}^2} \int_0^{z_i^{\rm in}}\frac{\Psi_i(z)}{\sum_j g_j\epsilon_j(\tilde{m}_jz)}zdz, \nonumber 
\label{Eq:intN_i}
\end{eqnarray}
where
\begin{eqnarray}
\Psi_i(z)&=&\int_{z}^\infty \frac{\psi_{s_i}(x)(x^2-z^2)}{e^x-(-1)^{2s_i}}dx  \\
&\simeq&a_\Psi\{ 1-\left[1+\exp(-b_\Psi\log_{10}z + c_\Psi) \right]^{-d_\Psi} \}, \nonumber
\end{eqnarray}
$z_i^{\rm in}=m_i/T_{\rm BH}^{\rm in}$ (the ratio of particle mass to initial BH temperature), $\tilde{m}_j = m_j/m_i$ and \cite{Cheek2022PRD}
\begin{eqnarray}
a_\Psi = 
\begin{cases}
2.457 & ({\rm scalar}) \\
0.897 & ({\rm fermion}) \\
0.2736 & ({\rm vector}) \\
0.0259 & ({\rm graviton}) \\
\end{cases},
\quad
b_\Psi = 
\begin{cases}
7.50218 & ({\rm scalar}) \\
12.3573 & ({\rm fermion}) \\
13.465 & ({\rm vector}) \\
22.325 & ({\rm graviton}) \\
\end{cases},
\end{eqnarray}
\begin{eqnarray}
c_\Psi = 
\begin{cases}
2.9437 & ({\rm scalar}) \\
8.7436 & ({\rm fermion}) \\
9.8134 & ({\rm vector}) \\
21.232 & ({\rm graviton}) \\
\end{cases},
\quad
d_\Psi = 
\begin{cases}
0.4208 & ({\rm scalar}) \\
0.3045 & ({\rm fermion}) \\
0.3049 & ({\rm vector}) \\
0.1207 & ({\rm graviton}) \\
\end{cases}.
\end{eqnarray}
%

%%----------------------------------------------------------------------------------
\subsection{Constraints from observations}
%%----------------------------------------------------------------------------------
Because the Hubble parameter $H(t)$ at time $t$ is less than or equal to the Hubble parameter during inflation, we obtain the lower limit of the initial mass of PBH as $M_{\rm BH}^{\rm in} \gtrsim 0.1$ g  (cosmic microwave background radiation (CMB) constraint) \cite{Fujita2014PRD}. In addition, because PBHs should be evaporated before big bang nucleosynthesis (BBN) \cite{Fujita2014PRD,Kohri1999PRD,Kawasaki2000PRD}, the upper limit $M_{\rm BH}^{\rm in} \lesssim 1 \times 10^9$ g is obtained (BBN constraint)\cite{Carr2021RPP}. Thus, we obtain $0.1 {\rm g} \le M_{\rm BH}^{\rm in} \le 10^9 {\rm g}$ ($4.6 \times 10^3 \lesssim M_{\rm BH}^{\rm in}/M_{\rm Pl} \lesssim 4.6 \times 10^{13}$) from the CMB and BBN constraints. In this paper, we assume that all PBHs evaporated completely during the radiation-dominant era in the next section.

%%----------------------------------------------------------------------------------
\section{Initial PBH density and DM mass spectrum \label{section:PBHandDM}}
%%----------------------------------------------------------------------------------

%%----------------------------------------------------------------------------------
\subsection{DM production by PBH}
%%----------------------------------------------------------------------------------
For simplicity, we assume that PBHs are the only source of DMs, and DMs only interact with the standard model particles gravitationally. Discussion of specific dark matter scenarios is out of topics in this study. We would like to perform a more detailed analysis for these specific scenarios in our future study.

The relic abundance of a DM particle species $i$ is typically expressed in terms of the density parameter as 
\begin{eqnarray}
\Omega_i= \frac{\rho_{i,0}}{\rho_{\rm c}} = \frac{m_i n_{i,0}}{\rho_{\rm c}},
\end{eqnarray}
where $\rho_{\rm c}$ denotes the critical density. $\rho_{i,0}$, $m_i$, and $n_{i,0}$ denote the today's energy density, mass, and today's number density of a DM particle species $i$, respectively. According to the entropy conservation
\begin{eqnarray}
\frac{n_{i,0}}{s_0} =\frac{n_i(T_{\rm ev})}{s(T_{\rm ev})},
\end{eqnarray}
we have
\begin{eqnarray}
\Omega_i = \frac{1}{\rho_{\rm c}}\frac{s_0}{s(T_{\rm ev})}m_i n_i(T_{\rm ev}) 
        = \frac{1}{\rho_{\rm c}} \frac{g_{*s}(T_0)T_0^3}{g_{*s}(T_{\rm ev})T_{\rm ev}^3}  m_i n_i(T_{\rm ev}).
\end{eqnarray}
Because the DM particle species $i$ is produced by PBH evaporation in our scenario, we obtain
\begin{eqnarray}
n_i(T_{\rm ev}) = \mathcal{N}_i n_{\rm BH}(T_{\rm ev}),
\end{eqnarray}
where $n_{\rm BH}(T_{\rm ev})$ denotes the PBH number density at the evaporation, which, for a monochromatic mass spectrum, can be related to the initial number density by
\begin{eqnarray}
n_{\rm BH}(T_{\rm ev})a(T_{\rm ev})^3 = n_{\rm BH}(T_{\rm in})a(T_{\rm in})^3.
\end{eqnarray}
Thus, we find
\begin{eqnarray}
\Omega_i =  \frac{1}{\rho_{\rm c}} \frac{g_{*s}(T_0)T_0^3}{g_{*s}(T_{\rm ev})T_{\rm ev}^3} \frac{a(T_{\rm in})^3}{a(T_{\rm ev})^3} \frac{\rho_{\rm BH}^{\rm in}}{M_{\rm BH}^{\rm in}} m_i \mathcal{N}_i,
\end{eqnarray}
where $\rho_{\rm BH}^{\rm in} = M_{\rm BH}^{\rm in}n_{\rm BH}(T_{\rm in})$.

Because we assume that all PBHs evaporated completely during the radiation-dominant era, the populations of PBHs remain a negligible component of the energy density of the Universe. In this case, the following simple form of the relic abundance is obtained \cite{Cheek2022PRD}
\begin{eqnarray}
\Omega_i = \frac{1}{\rho_{\rm c}} \frac{g_{*s}(T_0)T_0^3}{g_{*s}(T_{\rm in})T_{\rm in}^3} \frac{\rho_{\rm BH}^{\rm in}}{M_{\rm BH}^{\rm in}} m_i \mathcal{N}_i.\end{eqnarray}
Using Eqs. (\ref{Eq:PBHformationMin}) and (\ref{Eq:beta}), we obtain
\begin{eqnarray}
\Omega_ih^2 &\simeq&  1.571 \times 10^7  \left(\frac{\gamma}{0.2}\right)^{1/2} \left(\frac{106.75}{g_*(T_{\rm BH}^{\rm in})}\right)^{1/4}\left(\frac{M_{\rm Pl}}{M_{\rm BH}^{\rm in}}\right)^{3/2} \beta \left(\frac{m_i}{1{\rm GeV}}\right)\mathcal{N}_i\nonumber \\
&\simeq&  1.595 \left(\frac{\gamma}{0.2}\right)^{1/2} \left(\frac{106.75}{g_*(T_{\rm BH}^{\rm in})}\right)^{1/4}\left(\frac{1{\rm g}}{M_{\rm BH}^{\rm in}}\right)^{3/2} \beta \left(\frac{m_i}{1{\rm GeV}}\right)\mathcal{N}_i, 
\label{Eq:OmegaDM_PBH_i}
\end{eqnarray}
where $h$ denotes the dimension less Hubble parameter. 

The total DM density should be the sum of the relic abundance of the DM particle species $i$
\begin{eqnarray}
\Omega_{\rm DM}h^2 &=& \sum_{i=1}^{N_{\rm DM}} \Omega_i h^2,
\label{Eq:OmegaDM}
\end{eqnarray}
it should be consistent with the observed relic abundance of DM  \cite{Planck2020AA}
\begin{eqnarray}
\Omega_{\rm DM} h^2 = 0.11.
\end{eqnarray}
%

%%----------------------------------------------------------------------------------
\subsection{Single DM}
%%----------------------------------------------------------------------------------
First, we study the phenomenology of PBHs with a single DM species ($N_{\rm DM}=1$). 

Recall that PBHs are the only source of DMs, and DMs only interact with the standard model particles gravitationally. If there is a single DM particle species, the relic abundance of DM, $\Omega_{\rm DM}h^2$, should be controlled by mainly three parameters as shown in Eq.(\ref{Eq:OmegaDM_PBH_i}): the DM mass $m_{\rm DM}$, initial PBH density $\beta$, and initial PBH mass $M_{\rm BH}^{\rm in}$.

\begin{figure}[t]
\centering
\includegraphics[scale=1.0]{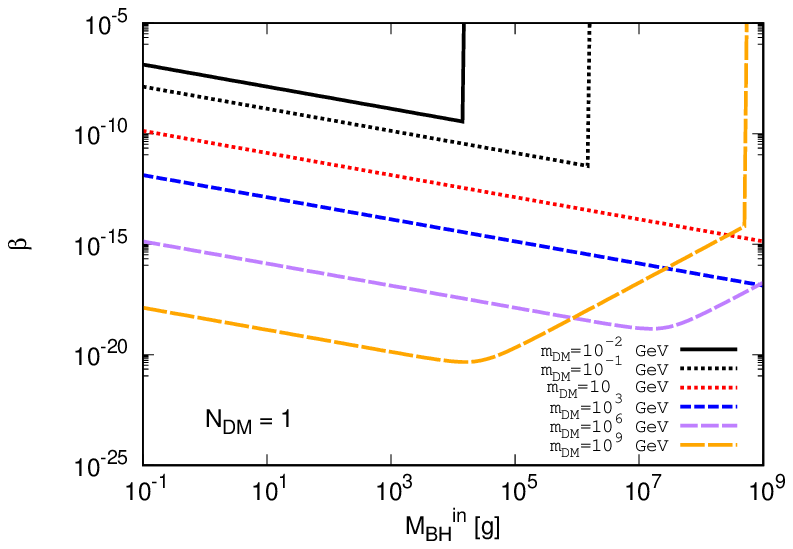}
\includegraphics[scale=1.0]{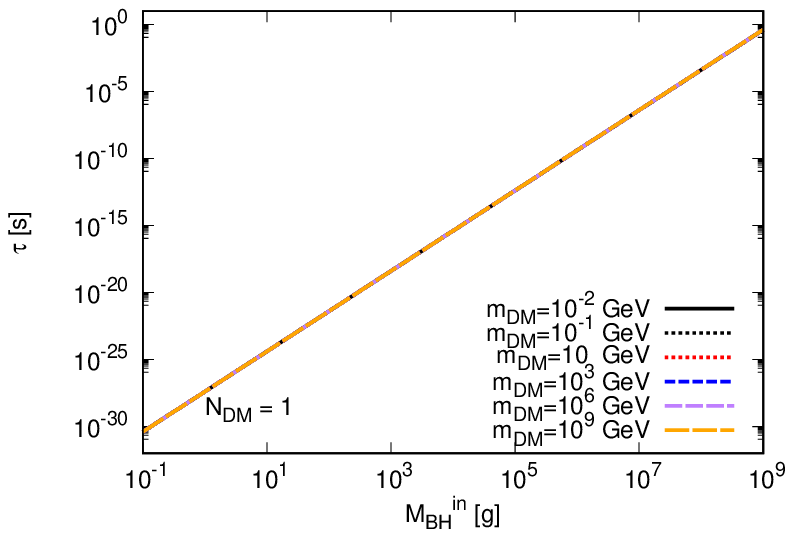}
\caption{Single DM species. DM is a scalar particle. Top panel: Allowed initial PBH density $\beta$ as a function of the initial PBH mass $M_{\rm BH}^{\rm in}$ for observed DM relic density $\Omega_{\rm DM} h^2 = 0.11$. Bottom panel: the PBH lifetime $\tau$ as a function of the initial PBH mass $M_{\rm BH}^{\rm in}$ for $m_{\rm DM} = 10^{-2}, \cdots, 10^9$ GeV. The six lines for $m_{\rm DM} = 10^{-2}, \cdots, 10^9$ GeV are undistinguished.}
\label{fig:single_dm_beta_tau_mbhin}
\end{figure}

Fig. \ref{fig:single_dm_beta_tau_mbhin} (top panel) shows the allowed initial PBH density $\beta$ as a function of the initial PBH mass $M_{\rm BH}^{\rm in}$ for observed DM relic density $\Omega_{\rm DM} h^2 = 0.11$, where we use the standard value of $g_*(T_{\rm BH}^{\rm in}) = 106.75$, and we assume that DMs are scalar particles. The observed DM abundance is consistent with a point on the curves for each DM mass. A point above the curves leads to an overproduction of DM for each DM mass. In contrast, a point below the curves leads to underproduction. The behavior of the curves in Fig. \ref{fig:single_dm_beta_tau_mbhin}  (top panel) is well understood, see for example \cite{Cheek2022PRD}. 

%--------------------------------------------------------------------
%-------------------------------------------------------------------

If the initial Hawking temperature $T_{\rm BH}^{\rm in}$ is higher than the DM mass $m_{\rm DM}$, PBHs begin producing DM particles immediately after formation and continue producing DMs during evaporation. Because $T_{\rm BH} \propto 1/M_{\rm BH}$, the relatively light initial PBHs can start producing DMs at an early stage, and the density of DM particles produced via PBH evaporation is linearly related to the initial density of PBHs. Heavier initial PBHs produce fewer heavy DM particles, and leading DM production can be started after crossing $T_{\rm BH} \simeq m_{\rm DM}$. Thus, some curves go up at a point for heavy DMs in Fig. \ref{fig:single_dm_beta_tau_mbhin}  (top panel) . 

Because we assume that PBHs have been evaporated entirely in the radiation-dominated era, the initial PBH density should be smaller than $\beta_c$. If the initial PBH density is larger than $\beta_c$, the relic abundance of DM does not depend on the initial PBH density \cite{Masina2020EPJP,Gondolo2020PRD,Bernal2021JCAP,Cheek2022PRD}. Thus, some curves are vertical lines at a point, $\beta = \beta_c$, in Fig. \ref{fig:single_dm_beta_tau_mbhin} (top panel) .

Fig. \ref{fig:single_dm_beta_tau_mbhin}  (top panel) shows that the heavier DM, up to approximately $10^6$ GeV, significantly contributes to the relic abundance and decreases the allowed initial PBH density in almost all ranges of $M_{\rm BH}^{\rm in} = 0.1 - 10^9$ g. For heavier DMs, such as $10^9$ GeV, the allowed initial PBH density increases with initial PBH mass for large PBH mass. From the viewpoint of future observations of PBHs, we are very interested in a case in which the initial PBH density is not very small. Moreover, the semiclassical approximation used by Hawking to derive the evaporation spectrum fails at the Planck scale. At the end of the Hawking evaporation, the horizon of  a black hole enters a physical region where quantum gravity cannot be neglected \cite{Dambrosio2021PRD,Soltani2021PRD}.

We set the range of DM mass to $10^{-2} \ {\rm GeV} \le m_{\rm DM}  \le 10^9 \ {\rm GeV}$ to reduce the cost of numerical calculations in this section. The lower mass $10^{-2} \ {\rm GeV}$ is used to avoid the warm DM consideration with PBHs in this study \cite{Fujita2014PRD,Masina2020EPJP,Cheek2022PRD}. The upper mass $10^9 \ {\rm GeV}$ is sufficient to show the existence of overlap of the curves in $\beta - M_{\rm BH}^{\rm in}$ plane. (This is the aim of this section). In the next section, we will extend the range of DM masses to more wide region.

Fig. \ref{fig:single_dm_beta_tau_mbhin} (bottom panel) shows the PBH lifetime $\tau$ as a function of the initial PBH mass $M_{\rm BH}^{\rm in}$ for $m_{\rm DM} = 10^{-2}, \cdots, 10^9$ GeV. The six lines for $m_{\rm DM} = 10^{-2}, \cdots, 10^9$ GeV are undistinguished, and the PBH lifetime is almost independent of DM mass. This independence of PBH lifetime may be understood by the following consideration. The evaporating PBHs predominantly radiate particles with mass less than the Hawking temperature. When the Hawking temperature rises above the  mass of the heavy particle, new radiation of the heavy particle can be started. The PBHs considered in this paper ($M_{\rm BH}^{\rm in} < 10^9$ g) have a high initial temperature ($T_{\rm in} > 10^5$ GeV). The PBH lifetime is determined by mainly all standard model particle spectra. Thus, the effect of heavy particles on the PBH lifetime may be small until the last stage of PBH evaporation, and the PBH lifetime is almost independent of DM mass. 

%%----------------------------------------------------------------------------------
\subsection{Degenerated DM mass spectrum}
%%----------------------------------------------------------------------------------
Next, we study the phenomenology of PBHs in the case of multiple DM particle species with degenerated DM mass spectrum:
\begin{eqnarray}
m_i = m_0, \quad (i=1,2,\cdots,N_{\rm DM}),
\label{Eq:dmMassSpectrumDegenerated}
\end{eqnarray}
where $m_i$ denotes the mass of a DM particle species in multiple DMs. All DM species have a common mass $m_0$.  For a complete analysis, an extremely large number of DM species such as $N_{\rm DM} \sim \infty$ should be considered; however, we would like to set the maximum number of DM particle species to 100 and consider four cases of $N_{\rm DM} = 1, 2, 10$, and $100$ to reduce the cost of numerical calculations in this section. 

%--------------------------------------------------------------------
\begin{figure}[t]
\centering
\includegraphics[scale=1.0]{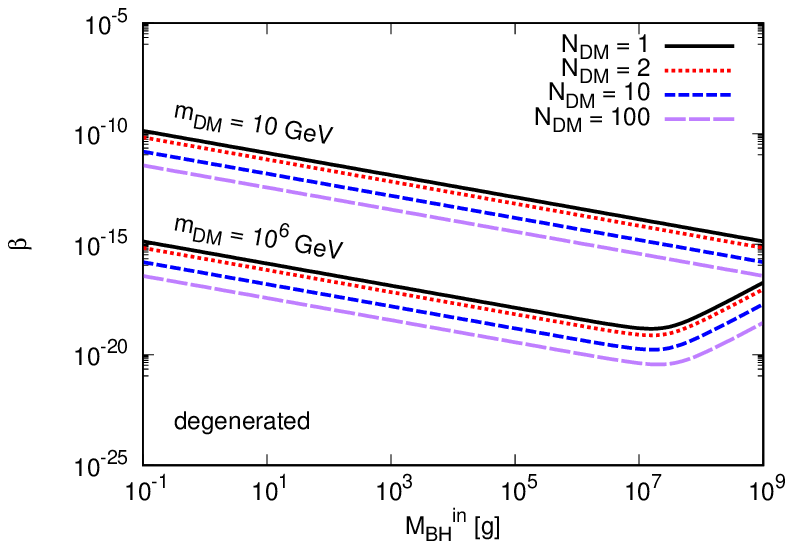}
\includegraphics[scale=1.0]{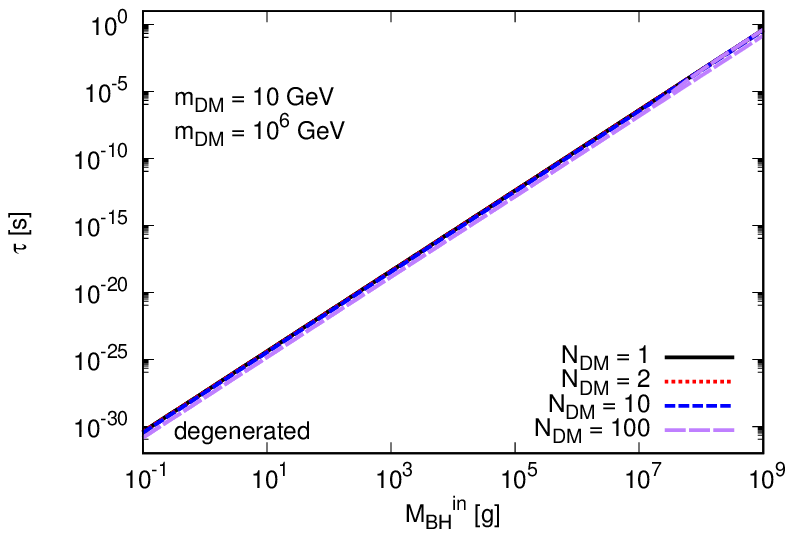}
\caption{Multiple DM species with degenerated mass spectrum. DMs are scalar particles. Top panel: Allowed initial PBH density $\beta$ as a function of the initial PBH mass $M_{\rm BH}^{\rm in}$ for observed DM relic density $\Omega_{\rm DM} h^2 = 0.11$. Bottom panel: the PBH lifetime $\tau$ as a function of the initial PBH mass $M_{\rm BH}^{\rm in}$ for $m_{\rm DM} = 10$ and $10^6$ GeV. The two lines for $m_{\rm DM} = 10^{-2}$ and $10^6$ GeV are undistinguished.}
\label{fig:degenerated_dm_beta_tau_mbhin}
\end{figure}
%-------------------------------------------------------------------

Fig. \ref{fig:degenerated_dm_beta_tau_mbhin} is similar to Fig. \ref{fig:single_dm_beta_tau_mbhin} but shows multiple DM particle species with degenerated DM mass spectrum. Again, we assume that all DMs are scalar particles. As benchmark cases, we show the curves for $m_{\rm DM} = 10$ GeV and $10^6$ GeV in the figure.

Fig. \ref{fig:degenerated_dm_beta_tau_mbhin} (top panel) shows that the allowed initial PBH density monotonically decreases with $N_{\rm DM}$. Because PBHs emit multiple particles simultaneously if these particle species have the same mass and spin, the total number of DM particles  produced by Hawking radiation monotonically increases with the number of DM species $N_{\rm DM}$ (the effect of a lifetime on the total number of DM particles produced is small, as we describe later). Thus, the allowed initial PBH density monotonically decreases with $N_{\rm DM}$. Because of this simple reduction mechanism of initial PBH density, each of the four curves for $N_{\rm DM} = 1, 2, 10$, and $100$ has the same shape. As we mentioned, see Fig. \ref{fig:single_dm_beta_tau_mbhin}  (top panel) for a single DM case, the heavier DM reduces the allowed initial PBH density for $m_{\rm DM} \lesssim 10^6$ GeV. Thus, the curves for $m_{\rm DM} = 10^6$ GeV are below the curves for $m_{\rm DM} = 10$ GeV in Fig. \ref{fig:degenerated_dm_beta_tau_mbhin} (top panel).

Fig. \ref{fig:degenerated_dm_beta_tau_mbhin} (bottom panel) shows that the PBH lifetime for $m_{\rm DM} = 10^{-2}$ and $10^6$ GeV are undistinguished. Because the PBH lifetime is almost independent of DM mass, see Fig. \ref{fig:single_dm_beta_tau_mbhin}  (bottom panel) for a single DM case, the two lines for $m_{\rm DM} = 10^{-2}$ and $10^6$ GeV almost overlap with each other.

As the evaporation rate increases with the number of degrees of particle contents, the PBH lifetime decreases with the additional DM particles. This short-lifetime effect is small, because a heavy particle, such as DM, can accelerate the PBH evaporation rate only in the final stage of the PBH's lifetime. The lifetime of the PBH for $N_{\rm DM} = 100$ is slightly longer than that for $N_{\rm DM} = 1, 2$, and $10$, and three lines for $N_{\rm DM} = 1, 2$, and $10$ almost overlap with each other.

We note that $\beta$-$M_{\rm BH}^{\rm in}$ curves for different $N_{\rm DM}$ differ from one another in the degenerated DM mass spectrum (top panel  in Fig. \ref{fig:degenerated_dm_beta_tau_mbhin}). As shown in the following subsection, this picture for the degenerated DM mass spectrum should be modified for the arithmetic sequence-like DM mass spectrum. 

%%----------------------------------------------------------------------------------
\subsection{Arithmetic sequence-like DM mass spectrum}
%%----------------------------------------------------------------------------------
We set the range of DM masses to
\begin{eqnarray}
m_0 \le m_i \le M, \quad  (i=1,2,\cdots,N_{\rm DM}),
\end{eqnarray}
and consider an arithmetic sequence like DM mass spectrum
\begin{eqnarray}
m_i = m_0 + (i-1)^p\frac{M-m_0}{(N_{\rm DM}-1)^p},
\label{Eq:dmMassSpectrumArithmetic}
\end{eqnarray}
for $N_{\rm DM} \neq 1$. For $N_{\rm DM} = 1$, we set $m_i = m_0$. Because the parameter $p$ controls the sparseness, or denseness, of the DM mass spectrum, as we show below, we consider $p$ as a sparseness parameter. 

In the case of $p = 0$, we obtain a degenerate DM mass $m_i = M$. In the case of $p=1$, we obtain an arithmetic sequence of DM masses 
\begin{eqnarray}
m_1 &=& m_0, \nonumber \\
m_2 &=& m_0 + \Delta m, \nonumber \\
m_3 &=& m_0 + 2\Delta m, \nonumber \\
 &\vdots& \nonumber \\
m_i &=& m_0 + (i-1)\Delta m, \nonumber \\
 &\vdots& \nonumber \\
m_{N_{\rm DM}} &=& M,
\end{eqnarray}
where
\begin{eqnarray}
\Delta m =  \frac{M - m_0}{ N_{\rm DM}-1},
\end{eqnarray}
for $N_{\rm DM} \neq 1$.  

%--------------------------------------------------------------------
\begin{figure}[t]
\centering
\includegraphics[scale=1.0]{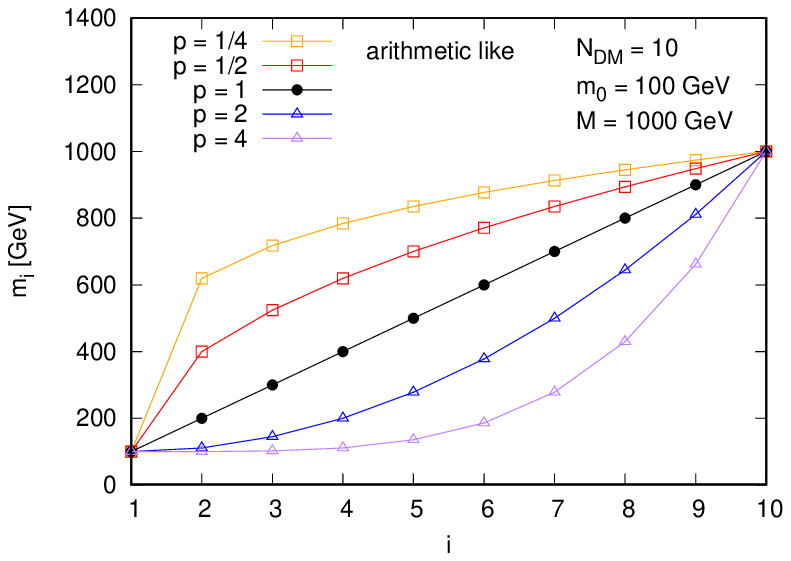}
\includegraphics[scale=1.0]{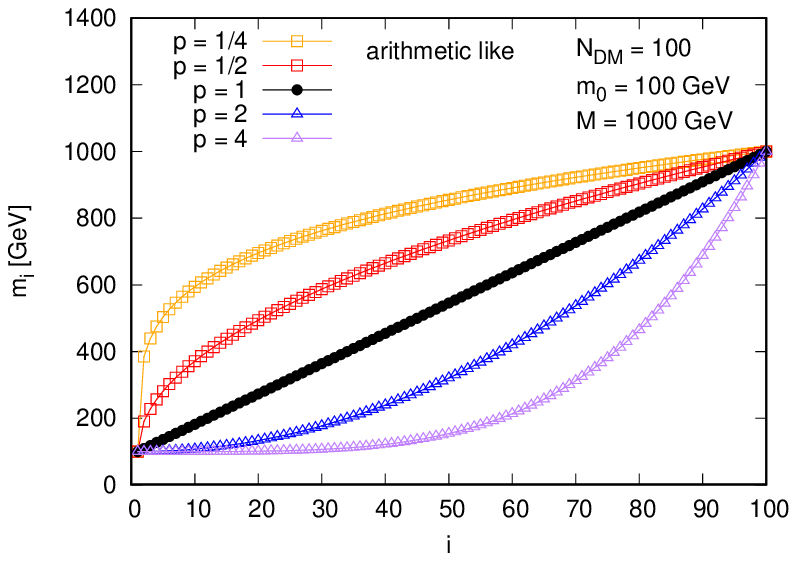}
\caption{Arithmetic sequence-like DM mass spectra for $N_{\rm DM}=10$ (top panel) and $100$ (bottom panel). DM mass spectra for $p=1/4, 1/2, 1, 2$, and $4$ with $m_0 = 100$ GeV and $M = 1000$ GeV are shown as benchmark cases.}
\label{fig:dm_mass_spectrum_arithmetic}
\end{figure}
%--------------------------------------------------------------------

%--------------------------------------------------------------------
\begin{figure}[t]
\centering
\includegraphics[scale=0.5]{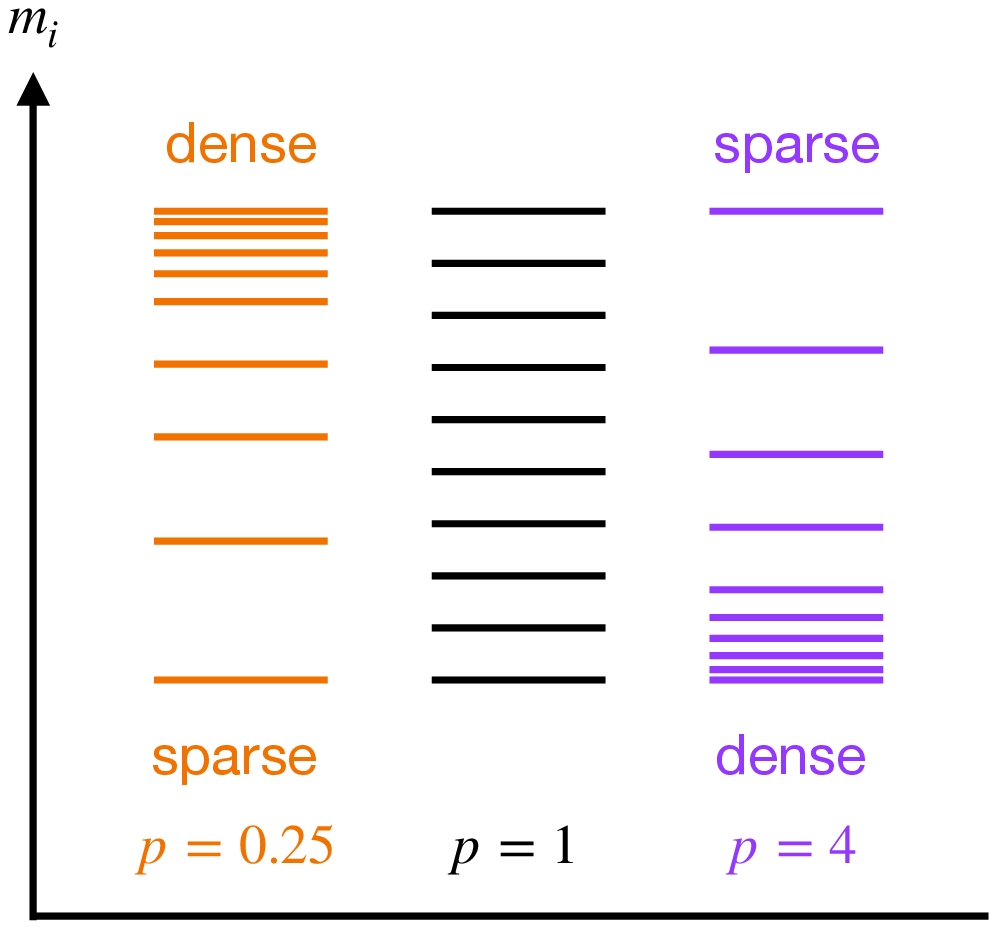}
\caption{Schematic image of arithmetic sequence-like DM mass spectra for $p = 0.25$ (sparse light DMs and dense heavy DMs), $p = 1$ (DMs with an arithmetic sequence mass spectrum), and $p = 4$ (sparse heavy DMs and dense light DMs.}
\label{fig:dm_mass_spectrum_arithmetic_schematic}
\end{figure}
%--------------------------------------------------------------------

%--------------------------------------------------------------------
\begin{figure}[t]
\centering
\includegraphics[scale=0.73]{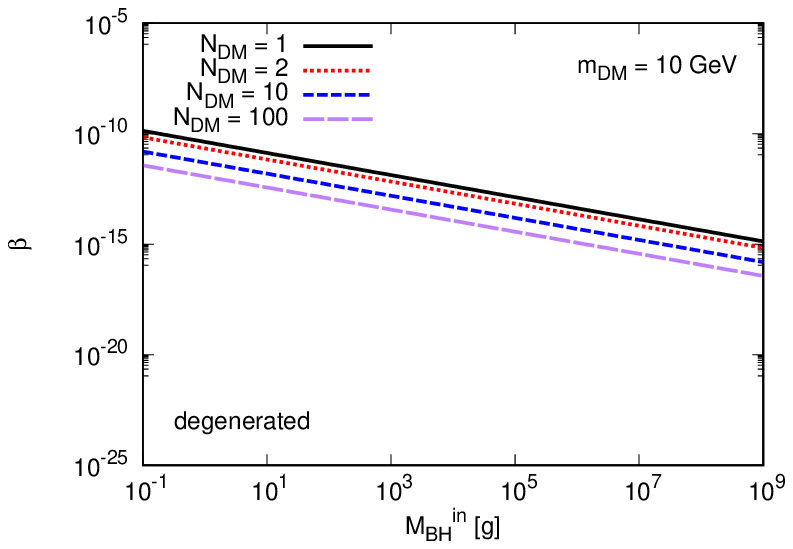}
\includegraphics[scale=0.73]{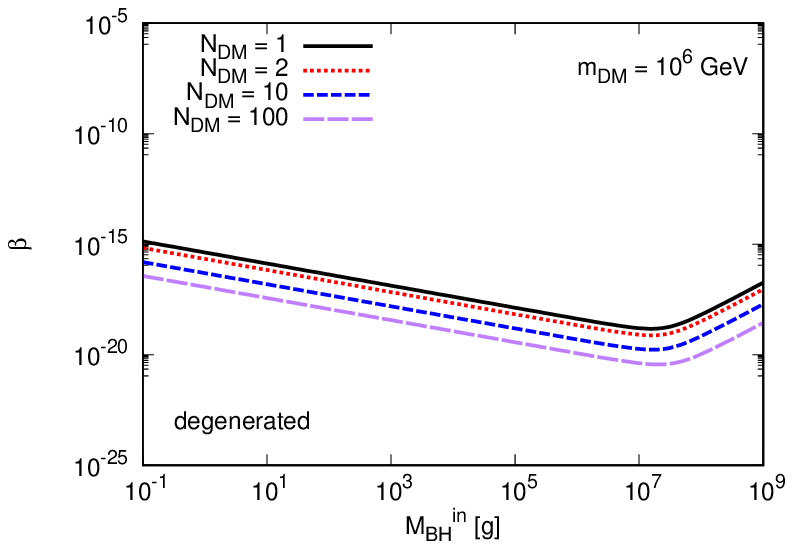}\\
\includegraphics[scale=0.73]{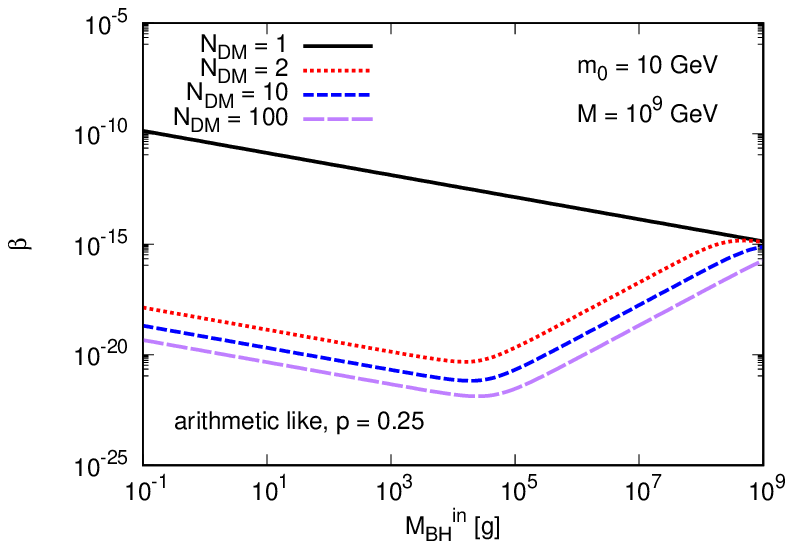}
\includegraphics[scale=0.73]{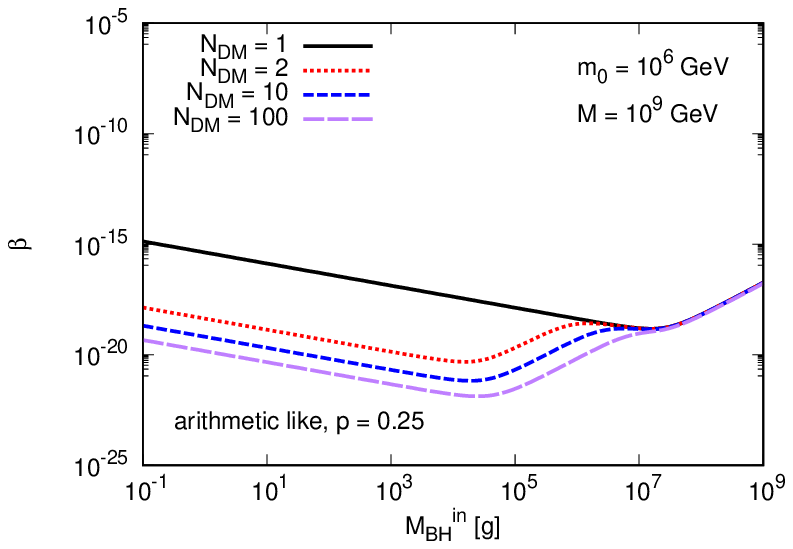}\\
\includegraphics[scale=0.73]{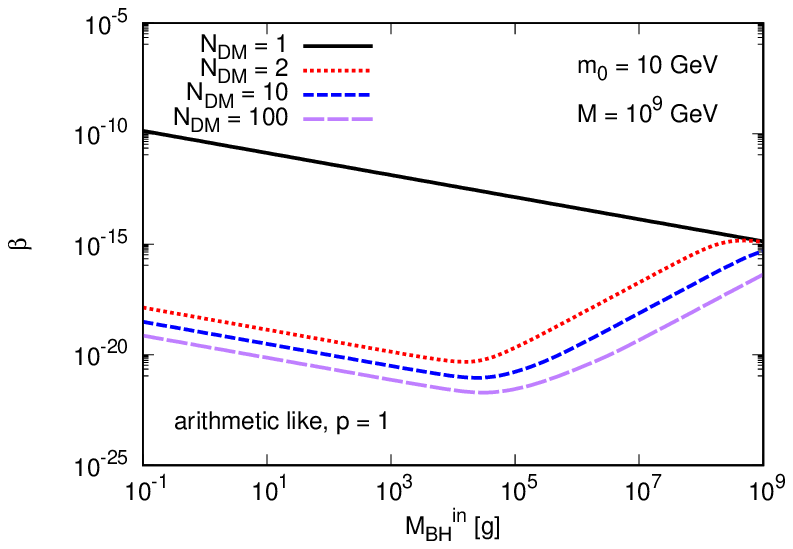}
\includegraphics[scale=0.73]{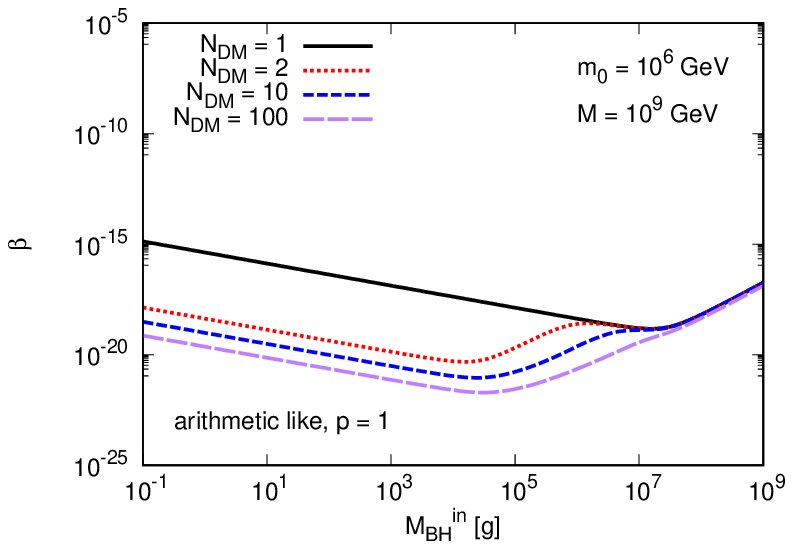}\\
\includegraphics[scale=0.73]{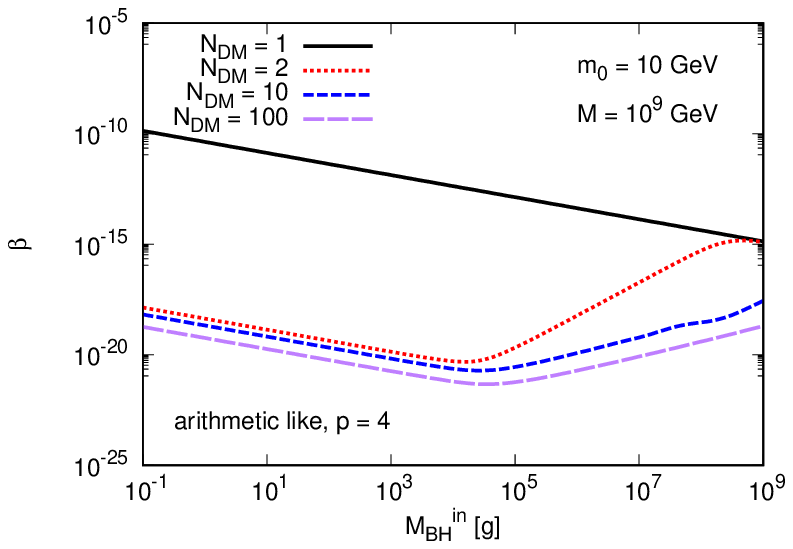}
\includegraphics[scale=0.73]{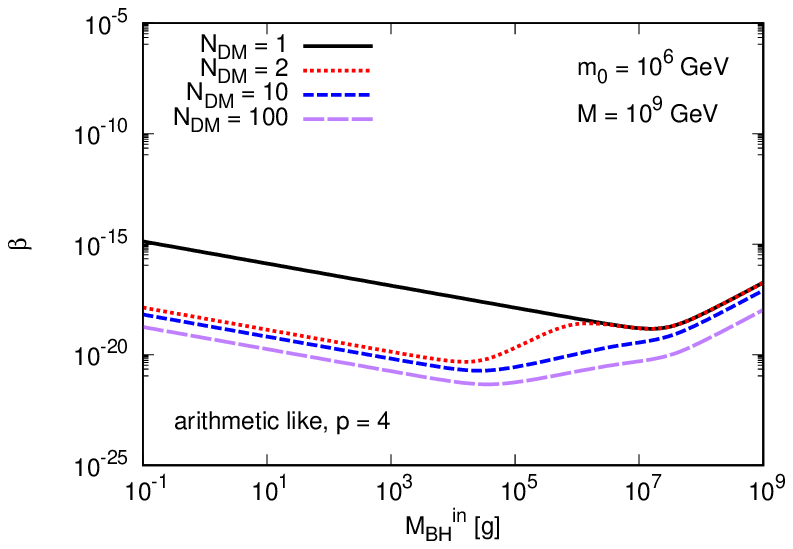}
\caption{Multiple DM species with an arithmetic mass spectrum. DMs are scalar particles. The top two panels show $\beta$-$M_{\rm BH}^{\rm in}$ curves in the degenerated DM mass spectrum case for comparison. Under the top two panels, the three left panels and right panels show the $\beta$-$M_{\rm BH}^{\rm in}$ curves for $10 \ {\rm GeV}  \le m_i \le 10^9 \  {\rm GeV}$ and $10^6 \ {\rm GeV}  \le m_i \le 10^9 \  {\rm GeV}$, respectively.}
\label{fig:arithmetic_dm_beta_mbhin}
\end{figure}
%-------------------------------------------------------------------
%--------------------------------------------------------------------
\begin{figure}[t]
\centering
\includegraphics[scale=0.73]{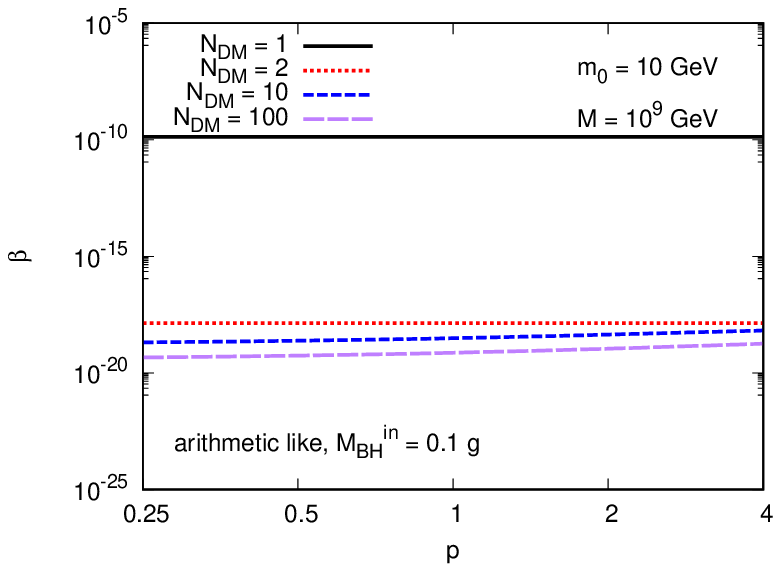}
\includegraphics[scale=0.73]{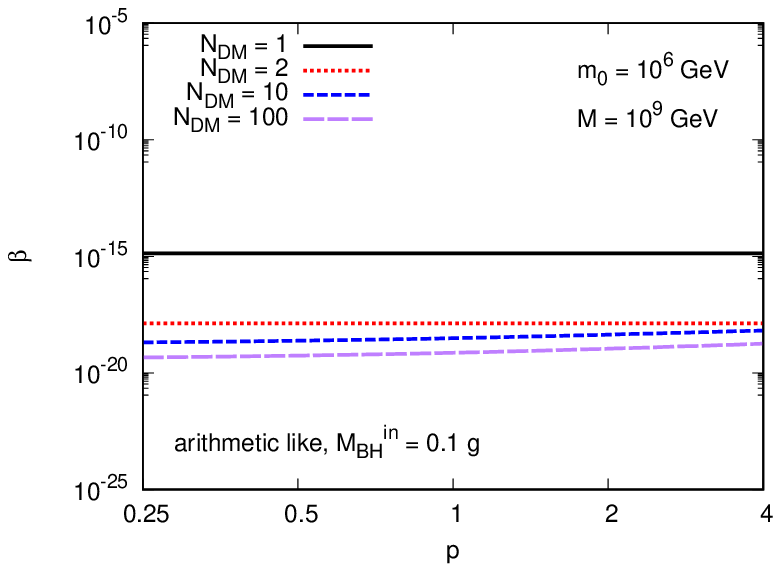}\\
\includegraphics[scale=0.73]{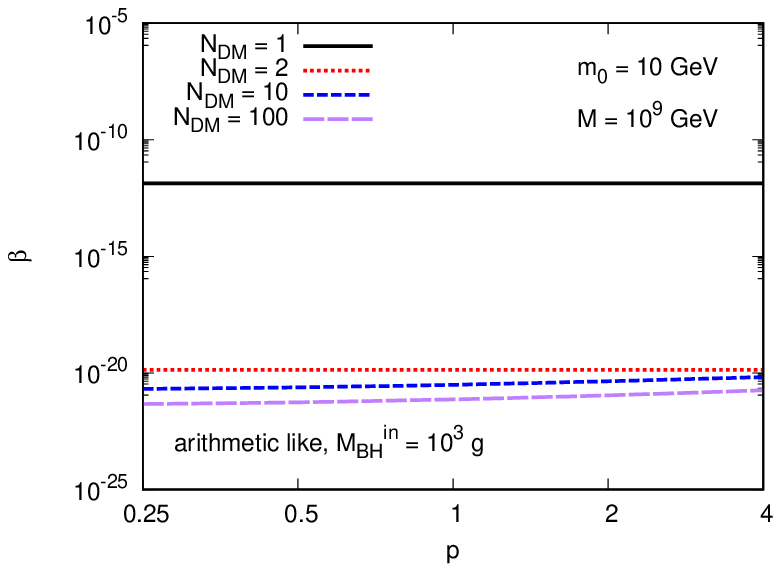}
\includegraphics[scale=0.73]{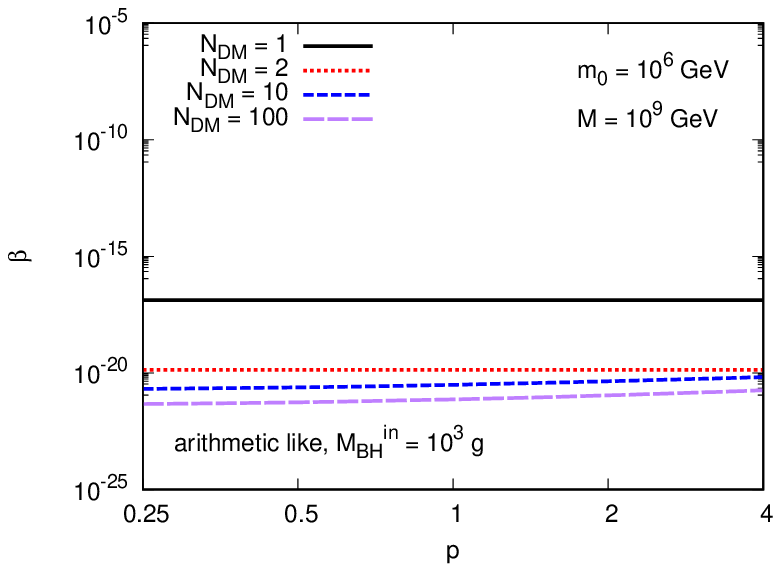}\\
\includegraphics[scale=0.73]{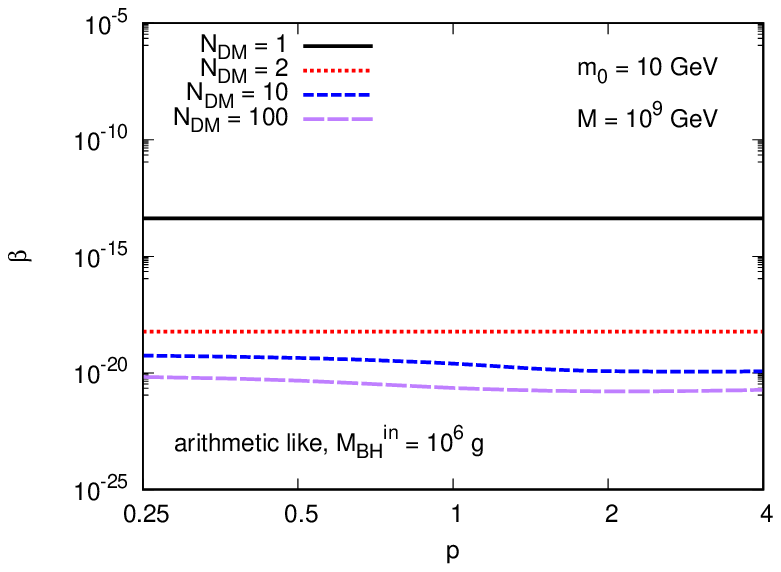}
\includegraphics[scale=0.73]{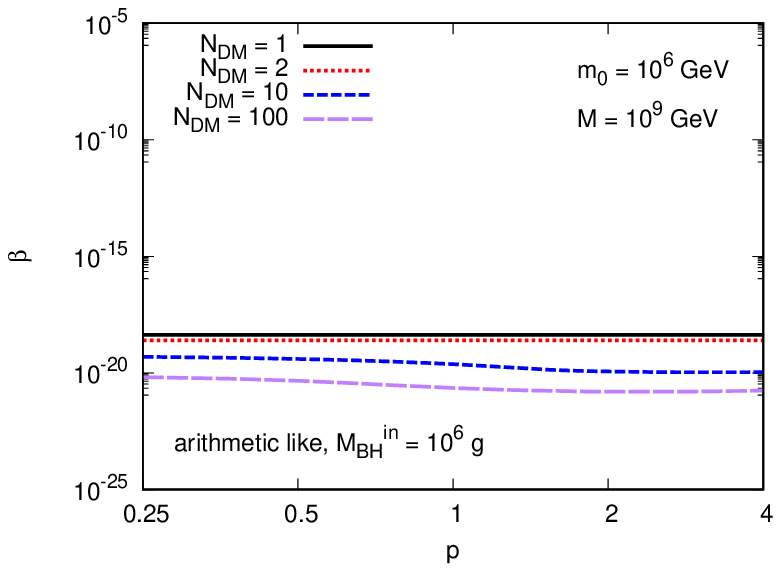}\\
\includegraphics[scale=0.73]{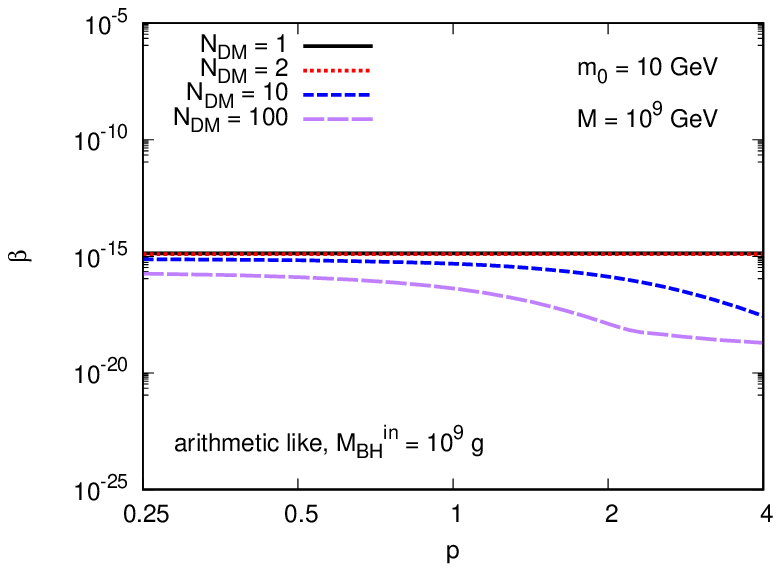}
\includegraphics[scale=0.73]{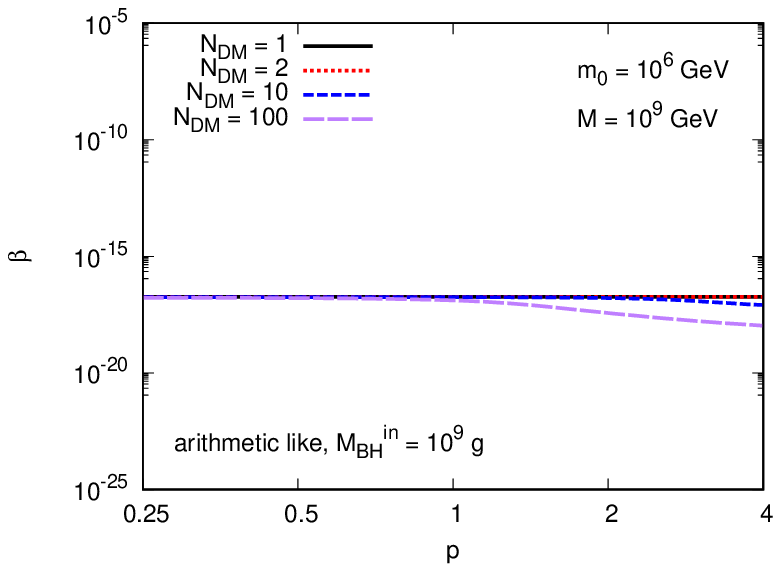}\\
\caption{Multiple DM species with an arithmetic mass spectrum. DMs are scalar particles. The four left panels and right panels show the $\beta$-$p$ curves for $10 \ {\rm GeV}  \le m_i \le 10^9 \  {\rm GeV}$ and $10^6 \ {\rm GeV}  \le m_i \le 10^9 \  {\rm GeV}$, respectively.}
\label{fig:arithmetic_dm_beta_p}
\end{figure}
%-------------------------------------------------------------------

For any value of sparseness parameter $p$ except for $p=0$ and $1$, an arithmetic sequence-like DM mass spectrum is obtained. Fig. \ref{fig:dm_mass_spectrum_arithmetic} shows an arithmetic sequence-like DM mass spectrum for $N_{\rm DM}=10$ (top panel) and $100$ (bottom panel) for $p=1/4, 1/2, 1, 2$, and $4$ with $m_0 = 100$ GeV and $M = 1000$ GeV as benchmark cases. As shown in Fig. \ref{fig:dm_mass_spectrum_arithmetic} (top panel), in the case of $p < 1$, the mass differences are sparse for the lighter DM and dense for the heavier DM. On the contrary,  in the case of $p > 1$, the DM mass spectrum becomes dense for the lighter DM and sparse for the heavier DM. Fig. \ref{fig:dm_mass_spectrum_arithmetic_schematic} depicts a schematic image of an arithmetic sequence-like DM mass spectrum for $p = 0.25$ (sparse light DMs and dense heavy DMs), $p = 1$ (DMs with an arithmetic sequence mass spectrum), and $p = 4$ (sparse heavy DMs and dense light DMs). Naturally, if the number of DM species is increased, the DM mass spectrum tends toward a continuous distribution, see Fig. \ref{fig:dm_mass_spectrum_arithmetic} (bottom panel).

Fig. \ref{fig:arithmetic_dm_beta_mbhin} shows the allowed initial PBH density $\beta$ as a function of the initial PBH mass $M_{\rm BH}^{\rm in}$ for multiple DMs with an arithmetic sequence-like mass spectrum for observed DM relic density $\Omega_{\rm DM} h^2 = 0.11$. We assume that DMs are scalar particles again. The top two panels show $\beta$-$M_{\rm BH}^{\rm in}$ curves in the degenerated DM mass spectrum case for comparison. Under the top two panels, the upper, middle, and bottom panels show $\beta$-$M_{\rm BH}^{\rm in}$ curves for $p = 0.25$, $p = 1$, and $p = 4$, respectively. 

In each of three left-bottom panels for $10 \ {\rm GeV}  \le m_i \le 10^9 \  {\rm GeV}$, the upper black solid curve shows the $\beta$-$M_{\rm BH}^{\rm in}$ relation for a single DM with mass $m_1 = m_0=10$ GeV. The three dotted or dashed curves under the black solid curve show the $\beta$-$M_{\rm BH}^{\rm in}$ relation for two species DMs ($N_{\rm DM}=2$) with masses $m_1 = m_0 = 10$ GeV and $m_2 = M = 10^9$ GeV for ten species DMs ($N_{\rm DM}=10$) with masses $m_i = 10, \cdots, 10^9$ GeV, and for a hundred species DMs ($N_{\rm DM}=100$) with masses $m_i = 10, \cdots, 10^9$ GeV, respectively. Because a heavy DM species with $m_i = 10^9$ GeV is included in all three cases for $N_{\rm DM} = 2, 10$, and $100$, the three dotted or dashed curves go up to a point around $M_{\rm BH}^{\rm in} = 10^4$ g, see Fig. \ref{fig:single_dm_beta_tau_mbhin} (top panel). The three right-bottom panels for $10^6 \ {\rm GeV}  \le m_i \le 10^9 \  {\rm GeV}$ are the same as the three left-bottom panels but for $m_0 = 10^6$ GeV.

In contrast to the degenerated DM mass spectrum, some curves tend to be close or overlap with each other in the case of the arithmetic DM mass spectrum. For example, in the left-bottom panel for $10 \ {\rm GeV}  \le m_i \le 10^9 \  {\rm GeV}$ and $p=4$, three dotted or dashed curves for $N_{\rm DM} = 2,10$, and $100$ for light initial PBH (left side in the panel) are closer than those for the heavy initial PBH (right side in the panel). 
The overlap of the curves is more obvious in the three right-bottom panels for $10^6 \ {\rm GeV}  \le m_i \le 10^9 \  {\rm GeV}$. For example, all four curves for $N_{\rm DM}=1,2,10$, and $100$ are indistinguishable for $p = 0.25$ and $M_{\rm BH}^{\rm in} \gtrsim 10^7$ g. For $p = 4$ and $M_{\rm BH}^{\rm in} \gtrsim 10^7$ g, the two curves for $N_{\rm DM}=1$ and $N_{\rm DM}=2$ overlap with each other. 

The cause of the overlapping of curves in Fig \ref{fig:arithmetic_dm_beta_mbhin} could be understated as follows. Because $\Omega_{\rm DM} h^2$ tends to increase with DM mass and $\beta \propto 1/ (\Omega_{\rm DM}h^2)$, the solid black curve for the lighter DM mass $m_0$ shows an upper limit of $\beta$. The three dotted or dashed curves for heavy DM go up to a point; however, these curves cannot exceed the same upper limit of $\beta$. Thus, some curves are overlapped with each other for relatively large $M_{\rm BH}^{\rm in}$. In addition, the large $m_0$ yields a  low upper limit of $\beta$, and some curves for $m_0=10^9$ GeV overlap over a larger area than the curves for $m_0=10$ GeV.

To illustrate these overlaps from a different perspective, we show the relation between $\beta$ and the sparseness parameter $p$ for multiple DMs with an arithmetic sequence-like mass spectrum for $\Omega_{\rm DM} h^2 = 0.11$ in Fig. \ref{fig:arithmetic_dm_beta_p}. The four left panels and right panels show the $\beta$-$p$ curves for $10 \ {\rm GeV}  \le m_i \le 10^9 \  {\rm GeV}$ and $10^6 \ {\rm GeV}  \le m_i \le 10^9 \  {\rm GeV}$, respectively. In Fig. \ref{fig:arithmetic_dm_beta_p}, the initial PBH mass $M_{\rm BH}^{\rm in}$ increases from the top panel to the bottom panel. In each panels, the horizontal axis shows the sparseness parameter $p$, the left side of each panel corresponds to the case of sparse light DMs and dense heavy DMs, and the right side of each panel  corresponds to the opposite case. The bottom-right panel in Fig. \ref{fig:arithmetic_dm_beta_p} shows that $\beta$-$p$ curves for different numbers of DM species $N_{\rm DM}$ overlap with each other for sparse light DMs and dense heavy DMs with heavy initial PBHs. 

The behavior of the curves in Fig. \ref{fig:arithmetic_dm_beta_p} can be understood as follows. Because a heavy initial PBH cannot emit a heavy DM particle species $i$ until the final stage of their lifetime and there are several heavy DMs in the case of small $p$, the total number of heavy DMs $\sum \mathcal{N}_i$ decreases with small $p$. From Eqs. (\ref{Eq:OmegaDM_PBH_i}) and (\ref{Eq:OmegaDM}), we obtain $\beta \propto 1/\sum \mathcal{N}_i$. Thus, as the initial PBH mass $M_{\rm BH}^{\rm in}$ becomes increasingly heavy, $\beta$-$p$ curves tend to be close to the upper limit of $\beta$ for small $p$. 

%%----------------------------------------------------------------------------------
\subsection{Geometric sequence-like DM mass spectrum}
%%----------------------------------------------------------------------------------

%--------------------------------------------------------------------
\begin{figure}[t]
\centering
\includegraphics[scale=1.0]{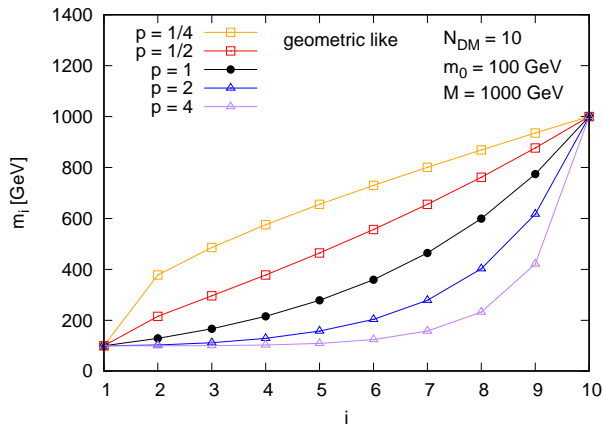}
\caption{Geometric sequence-like DM mass spectrum for $N_{\rm DM}=10$. DM mass spectra for $p=1/4, 1/2, 1, 2$, and $4$ with $m_0 = 100$ GeV and $M = 1000$ GeV are shown as benchmark cases.}
\label{fig:dm_mass_spectrum_geometric}
\end{figure}
%--------------------------------------------------------------------

%--------------------------------------------------------------------
\begin{figure*}[t]
\centering
\includegraphics[scale=0.8]{fig_a_m1e-1_M1e9_m01e6_b_m.eps}
\includegraphics[scale=0.8]{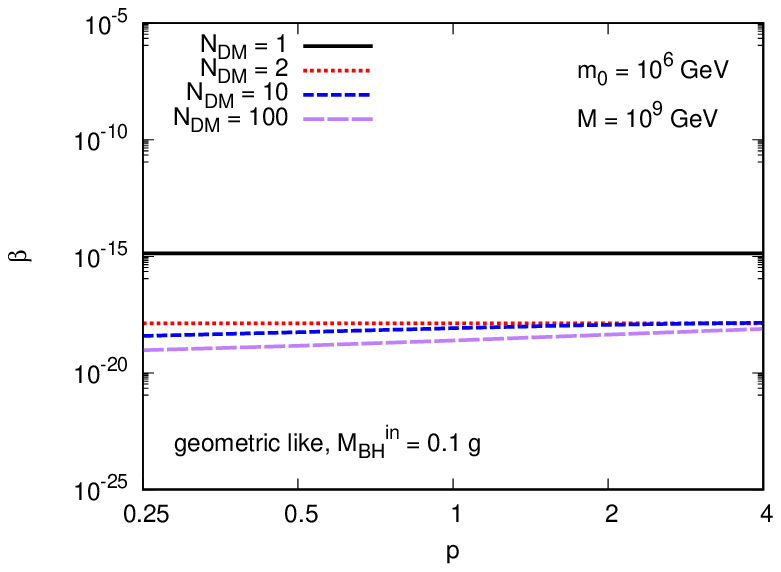}\\
\includegraphics[scale=0.8]{fig_a_m1e9_M1e9_m01e6_b_m.eps}
\includegraphics[scale=0.8]{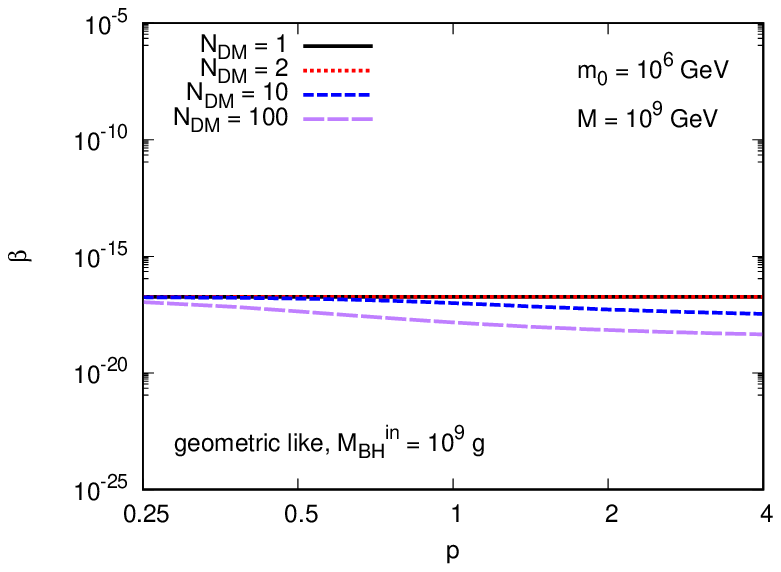}
\caption{$\beta$-$p$ curves for $M_{\rm BH}^{\rm in} = 0.1$ g and $M_{\rm BH}^{\rm in} = 10^9$ g. The two left panels show $\beta$-$p$ curves in the arithmetic DM mass spectrum case. The two right panels show $\beta$-$p$ curves in the geometric DM mass spectrum case.}
\label{fig:geometric_dm_beta_p}
\end{figure*}
%-------------------------------------------------------------------
Next, we hold the range of DM masses $m_0 \le m_i \le M$ $(i=1,2,\cdots,N_{\rm DM})$, but consider a geometric sequence-like DM mass spectrum
\begin{eqnarray}
m_i = m_0 \left(\frac{M}{m_0}\right)^{(i-1)^p/(N_{\rm DM}-1)^p}, 
\label{Eq:dmMassSpectrumGometric}
\end{eqnarray}
for $N_{\rm DM} \neq 1$. For $N_{\rm DM} = 1$, we set $m_i = m_0$.

In the case of $p = 0$, we obtain a degenerate DM mass $m_i = M$.  In the case of $p=1$, we obtain a geometric sequence of DM masses 
\begin{eqnarray}
m_1 &=& m_0, \nonumber \\
m_2 &=& m_0 r, \nonumber \\
m_3 &=& m_0 r^2, \nonumber \\
 &\vdots& \nonumber \\
m_i &=& m_0 r^{i-1},\nonumber \\
 &\vdots& \nonumber \\
m_{N_{\rm DM}} &=& M,
\end{eqnarray}
where
\begin{eqnarray}
r = \left(\frac{M}{m_0}\right)^{1/(N_{\rm DM}-1)}, 
\end{eqnarray}
for $N_{\rm DM} \neq 1$.

For any value of sparseness parameter $p$ except for $p=0$ and $1$, a geometric sequence-like DM mass spectrum is obtained. In Fig. \ref{fig:dm_mass_spectrum_geometric}, the geometric sequence-like DM mass spectrum for $N_{\rm DM}=10$, $p=1/4, 1/2, 1, 2$, and $4$, $m_0 = 100$ GeV and $M = 1000$ GeV are shown as benchmark cases. Similar to the arithmetic DM mass spectrum, there are sparse light DMs and dense heavy DMs in the case of small $p$. In contrast, in the case of large $p$, there are sparse heavy DMs and dense light DMs. 

Because of this similarity between the arithmetic sequence-like DM mass spectrum in Eq.(\ref{Eq:dmMassSpectrumArithmetic}) and the geometric sequence-like DM mass spectrum in Eq.(\ref{Eq:dmMassSpectrumGometric}), we have obtained $\beta$-$M_{\rm BH}^{\rm in}$ and $\beta$-$p$ curves similar to Fig. \ref{fig:arithmetic_dm_beta_mbhin} and Fig. \ref{fig:arithmetic_dm_beta_p}, respectively, for the geometric sequence-like DM mass spectrum. We would like to avoid showing all the results; however, we show four sample panels to demonstrate this similarity in Fig. \ref{fig:geometric_dm_beta_p}. Fig. \ref{fig:geometric_dm_beta_p} shows the $\beta$-$p$ curves for $M_{\rm BH}^{\rm in} = 0.1$ g (top panels) and $M_{\rm BH}^{\rm in} = 10^9$ g (bottom panels) as benchmark cases. The two left panels show $\beta$-$p$ curves in the arithmetic sequence-like DM mass spectrum case for comparison. The two right panels show $\beta$-$p$ curves in the geometric sequence-like DM mass spectrum case. Because the sparseness of the geometric sequence-like DM mass spectrum for a fixed value of $p$ is different from that of the arithmetic sequence-like DM mass spectrum for the same value of $p$, the shapes of their curves are different; however, the basic behaviors of the curves in the two left and right panels are the same. 

Thus, we finally conclude that, regardless of the detailed structure of the mass spectra of multiple DMs, such as arithmetic sequence-like or geometric sequence-like mass spectrum, $\beta$-$M_{\rm BH}^{\rm in}$ and $\beta$-$p$ curves for different numbers of DM species $N_{\rm DM}$ may tend to overlap with each other for heavy initial PBHs. 

%%----------------------------------------------------------------------------------
\subsection{DM spin}
%%----------------------------------------------------------------------------------
We would like to comment on the DM spin. We have assumed that all DM species are scalar particles ($s=0$). According to Cheek et al., the initial PBH densities required to produce the observed relic abundance of DM depend on the DM spin, varying in $\sim 2$ orders of magnitude between a spin-2 and a scalar DM in the case of Schwarzschild PBH \cite{Cheek2022PRD}. Although this variation of $\sim 2$ orders of magnitude may have a strong effect in some situations, the DM spin effect does not make a significant difference in the conclusion of this paper. Thus, we have shown only the case of scalar DMs in this section.

%%----------------------------------------------------------------------------------
\section{Constraint on DM masses in four categories \label{sec:four_cases}}
%%----------------------------------------------------------------------------------
Up to now, we have focused on the DM for $10^{-2}$ GeV $\le m_{\rm DM} \le 10^{9}$ GeV. This mass range is sufficient to show the existence of overlap of the curves in $\beta - M_{\rm BH}^{\rm in}$ plane. (The previous section showed the existence of this overlap). 

In this section, we discuss the other cases with different DM masses and show the constraint on the masses of multiple DM by existing observations. For a single DM species, the limits on DM mass providing the correct relic abundance in the four categories (the categories will be defined soon) have been estimated by Lennon, et al. \cite{Lennon2018JCAP}. Meanwhile, this study considers the constraint on the multiple DM masses in the same four categories by using the methods in Ref. \cite{Lennon2018JCAP}.

%%----------------------------------------------------------------------------------
\subsection{Single DM species}
%%----------------------------------------------------------------------------------
According to Lennon, et al., \cite{Lennon2018JCAP}, we consider the `light'  and `heavy' DM cases. For the `light' DM, the initial temperature of PBH $T_{\rm in}$ is greater than all mass scales. On the other hand, $T_{\rm in}$ is below the DM mass in the `heavy' DM case. The `slow' and `fast' decay cases have also been considered. $B(t)\equiv t_{\rm H}/\tau_{\rm dec}$ was defined to introduce two categories of PBH decay time scale. Here, $t_{\rm H}(t)$ is the Hubble time and $\tau_{\rm dec}(M_{\rm BH}(t))$ is the characteristic PBH decay time. The `slow' decay case is defined as $B_{\rm in} \ll 1$, where $B_{\rm in}$ is the initial value of $B$, and `fast' decay case is defined as $B_{\rm in} \gg 1$. 

This paper now has four categories: (1) light and slow, (2) light and fast, (3) heavy and slow, and (4) heavy and fast. Moreover, good analytic solutions for these four categories are obtained for a single DM species $\chi$ \cite{Lennon2018JCAP}.

{\bf Light DM and slow decay:} In the case of light DM and slow decay, the temperature of the standard model radiation bath at the end of PBH decay, the reheat temperature $T_{\rm RH}$, and the yield of DM $\chi$, $Y_\chi=n_\chi/s$ are
\begin{eqnarray}
T_{\rm RH, \chi}^{\rm slow} \simeq 1.09 \frac{e_{\rm T, \chi}^{1/2}}{g_\ast^{1/4}}M_{\rm Pl}\left( \frac{M_{\rm Pl}}{M_{\rm BH}^{\rm in}} \right)^{3/2},
\end{eqnarray}
and 
\begin{eqnarray}
Y_{\chi}^{\rm slow} \simeq 0.49 \frac{f_\chi g_\chi}{g_\ast^{1/4}e_{\rm T, \chi}^{1/2}}  \left( \frac{M_{\rm Pl}}{M_{\rm BH}^{\rm in}} \right)^{1/2},
\label{Eq:Y_chi_slow}
\end{eqnarray}
respectively, where $g_\ast$ is the effective number of degrees of freedom of standard model particles at $T_{\rm RH}$, and
\begin{eqnarray}
e_{\rm T, \chi} = e_{\rm T}^{\rm SM} + e_{\rm T}^\chi,
\end{eqnarray}
is the total emission coefficient for standard model particles, $e_{\rm T}^{\rm SM} \simeq 4.38\times 10^{-3}$, and single DM species $\chi$, $e_{\rm T}^\chi = e_\chi g_\chi$. The numerical value of $e_\chi$ and $f_\chi$ are given by
\begin{eqnarray}
e_\chi = 
\begin{cases}
7.24 \times 10^{-5} & (s=0) \\
4.09 \times 10^{-5}  & (s=1/2) \\
1.68 \times 10^{-5}  & (s=1) \\
 5.5\times 10^{-6}  & (s=3/2) \\
1.92 \times 10^{-6}  & (s=2) \\
\end{cases},
\quad
f_\chi = 
\begin{cases}
6.66 \times 10^{-4} & (s=0) \\
2.43 \times 10^{-4}  & (s=1/2) \\
7.40 \times 10^{-5}  & (s=1) \\
2.1\times 10^{-5}  & (s=3/2) \\
5.53 \times 10^{-6}  & (s=2) \\
\end{cases}.
\end{eqnarray}

{\bf Light DM and fast decay:} For the light DM and fast decay cases, the reheat temperature and the yield of DM $\chi$ are given by
\begin{eqnarray}
T_{\rm RH, \chi}^{\rm fast} \simeq \left( \frac{30 n_{\rm BH}(T_{\rm in}) M_{\rm BH}^{\rm in}}{\pi^2 g_\ast} \right)^{1/4},
\end{eqnarray}
and 
\begin{eqnarray}
Y_{\chi}^{\rm fast} \simeq 0.50 \frac{f_\chi g_\chi}{g_\ast^{1/4}e_{\rm T, \chi}} \left( \frac{n_{\rm BH}(T_{\rm in})}{M_{\rm Pl}^3} \right)^{1/4} \left( \frac{M_{\rm BH}^{\rm in}}{M_{\rm Pl}} \right)^{5/4},
\end{eqnarray}
respectively.

{\bf Heavy DM:} For the heavy DM case, the reheat temperature is similar to that of the light DM case. The yields of DM is
\begin{eqnarray}
Y_{\rm heavy} = Y_{\rm light} \frac{T^2_{\rm in}}{m_\chi^2}d_s^2,
\end{eqnarray}
where $d_s$ is estimated as $3.2$ and $3.6$ for bosons and fermions, respectively.

{\bf Limits on DM mass:} According to Lennon, et al., \cite{Lennon2018JCAP}, we require that the relations
\begin{eqnarray}
T_{\rm RH} > 3 \ {\rm MeV},
\end{eqnarray}
and
\begin{eqnarray}
m_\chi Y_\chi = 0.42 \ {\rm eV},
\end{eqnarray}
should be satisfied. The first and second requirement comes from BBN \cite{Kawasaki2000PRD} and $\Omega_{\rm DM} h^2=0.11$, respectively. Without free-streaming constraint in the light case, the limit on DM mass for the single DM species is obtained as shown in Table.\ref{tab:limit_mchi} \cite{Lennon2018JCAP}.

%=============================================
\begin{table}[t]
\caption{Limits on DM mass [GeV] for the correct relic abundance for the single DM case \cite{Lennon2018JCAP}.}
\small
\begin{center}
\begin{tabular}{|c|cccc|}
\hline
$s$  & Light and slow & Light and fast & Heavy and slow & Heavy and fast\\
\hline
0 &  $[2.6 \times 10^{-7}, 0.8]$  & $[3.1 \times 10^{-7}, 2.8 \times 10^{13}]$ &  $[3.4 \times 10^{9}, M_{\rm Pl}]$ &  $[2.9 \times 10^{9}, M_{\rm Pl}]$\\
1/2 &  $[3.6 \times 10^{-7}, 1.1]$  & $[4.2 \times 10^{-7}, 3.9 \times 10^{13}]$ &  $[3.1 \times 10^{9}, M_{\rm Pl}]$ &  $[2.6 \times 10^{9}, M_{\rm Pl}]$\\
1 &  $[7.8 \times 10^{-7}, 2.4]$  & $[9.2 \times 10^{-7}, 8.5 \times 10^{13}]$ &  $[1.1 \times 10^{9}, M_{\rm Pl}]$ &  $[9.6 \times 10^{8}, M_{\rm Pl}]$\\
2/3 &  $[2 \times 10^{-6}, 6]$  & $[2 \times 10^{-6}, 2 \times 10^{14}]$ &  $[5 \times 10^{8}, M_{\rm Pl}]$ &  $[5 \times 10^{8}, M_{\rm Pl}]$\\
2 &  $[6.3 \times 10^{-6}, 19]$  & $[7.4 \times 10^{-6}, 6.8 \times 10^{14}]$ &  $[1.4 \times 10^{8}, M_{\rm Pl}]$ &  $[1.2 \times 10^{8}, M_{\rm Pl}]$\\
\hline
\end{tabular}
\end{center}
\label{tab:limit_mchi}
\end{table}
%=============================================

Lennon et al. discussed the initial DM velocities and their subsequent redshift when DM was created by the evaporation of PBHs \cite{Lennon2018JCAP}. Consequently, the light DM mass case was found to be significantly constrained by the free streaming of the ultra-relativistically emitted DM particles and the almost the entire parameter space for light DM is excluded. However, this paper also includes the light DM case in the numerical calculations, as presented in section 3 in Ref. \cite{Lennon2018JCAP}.

%%----------------------------------------------------------------------------------
\subsection{Multiple DM species}
%%----------------------------------------------------------------------------------
In the multiple DM species cases, the method in the previous subsection should be extended with the relations
\begin{eqnarray}
m_\chi \quad \rightarrow  \quad m_1+ m_2 + m_3+ \cdots + m_{N_{\rm DM}}
\end{eqnarray}
and
\begin{eqnarray}
e_{\rm T,\chi} = e_{\rm T}^{\rm SM} + e_{\rm T}^\chi  \quad  \rightarrow  \quad  e_{\rm T} = e_{\rm T}^{\rm SM} + \sum_{i=1}^{N_{\rm DM}} e_i g_i.
\end{eqnarray}
The required condition then becomes
\begin{eqnarray}
m_\chi Y_\chi = 0.42 \ {\rm eV} \quad \rightarrow \quad \sum_{i=1}^{N_{\rm DM}} m_i Y_i  =  0.42 \ {\rm eV} \equiv \tilde{m}_{\rm DM} Y_\chi,
\end{eqnarray}
where an effective DM mass $\tilde{m}_{\rm DM}$ was defined to compare the result of the multiple and single DM cases. 

For simplicity, $f_i=f_\chi$ and $g_i = g_\chi$ were assumed. In this case, $e_{\rm T} = e_{\rm T}^{\rm SM} + e_{\rm T}^\chi N_{\rm DM}$ was obtained and the effective DM masses can be estimated as
\begin{eqnarray}
\tilde{m}_{\rm DM}=
\begin{cases}
 \left( \frac{e_{\rm T, \chi}}{e_{\rm T}} \right)^{1/2}\sum_{i=1}^{N_{\rm DM}} m_i  & (\rm light, slow), \\
 \frac{e_{\rm T, \chi}}{e_{\rm T}}\sum_{i=1}^{N_{\rm DM}} m_i  &(\rm light, fast), \\
 \left( \frac{e_{\rm T}}{e_{\rm T, \chi}} \right)^{1/2} \left(\sum_{i=1}^{N_{\rm DM}} \frac{1}{m_i}\right)^{-1} &(\rm heavy, slow), \\
  \frac{e_{\rm T}}{e_{\rm T, \chi}} \left( \sum_{i=1}^{N_{\rm DM}} \frac{1}{m_i} \right)^{-1} & (\rm heavy, fast).
\end{cases}
\label{Eq:tilde_mdm}
\end{eqnarray}
%

%%----------------------------------------------------------------------------------
\subsection{Degenerated DM mass spectrum}
%%----------------------------------------------------------------------------------
In the degenerated DM mass spectrum case, all of DM have the common mass $m_0$. We obtain 
\begin{eqnarray}
\tilde{m}_{\rm DM}=
\begin{cases}
\left( \frac{e_{\rm T, \chi}}{e_{\rm T}} \right)^{1/2}m_0N_{\rm DM} & (\rm light, slow),\\
\frac{e_{\rm T, \chi}}{e_{\rm T}}m_0N_{\rm DM}  & (\rm light, fast), \\
 \left( \frac{e_{\rm T}}{e_{\rm T, \chi}} \right)^{1/2} \frac{m_0}{N_{\rm DM}} & (\rm heavy, slow), \\
\frac{e_{\rm T}}{e_{\rm T, \chi}}  \frac{m_0}{N_{\rm DM}} & (\rm heavy, fast),
\end{cases}
\end{eqnarray}
and require that the relations
\begin{eqnarray}
m_\chi^{\rm min} <  m_0 < m_\chi^{\rm max}, \quad
m_\chi^{\rm min} <  \tilde{m}_{\rm DM} < m_\chi^{\rm max},
\end{eqnarray}
should be satisfied, where $m_\chi^{\rm min}$ and $m_\chi^{\rm max}$ are the lower and upper limits of the DM mass in Table.\ref{tab:limit_mchi}, respectively.

For example, for $N_{\rm DM}=3$ and $s=0$, the effective DM mass is related to the common mass as follows
\begin{eqnarray}
\tilde{m}_{\rm DM}=
\begin{cases}
0.376 m_0 & (\rm light, slow), \\
0.0472 m_0  & (\rm light, fast), \\
2.66 m_0 & (\rm heavy, slow), \\
21.2 m_0 & (\rm heavy, fast),
\end{cases}
\end{eqnarray}
and the allowed region of DM mass to be
\begin{eqnarray}
\tilde{m}_{\rm DM}[{\rm GeV}] &\simeq&
\begin{cases}
2.6\times 10^{-7}  - 0.3, & (\rm light, slow), \\
3.1\times 10^{-7}  - 1.3 \times 10^{13}, & (\rm light, fast), \\
9.0 \times 10^9  - 1.2 \times 10^{19}, &  (\rm heavy, slow),\\
6.1 \times 10^{10}  - 1.2 \times 10^{19}, &  (\rm heavy, fast).
\end{cases}
\end{eqnarray}
It has been observed that for the light DM case, the upper limits of the DM mass decreased from $0.8 \ {\rm GeV}$ to $0.3 \ {\rm GeV} $ (light, slow) and $2.8\times 10^{13} \ {\rm GeV}$ to $1.3 \times 10^{13}\ {\rm GeV} $ (light, fast).  For the heavy DM case, the lower limits of DM mass increased from $3.4 \times 10^9\ {\rm GeV}$ to $9.0 \times 10^{9}\ {\rm GeV} $ (heavy, slow) and $2.9 \times 10^9\  {\rm GeV}$ to $6.1 \times 10^{10}\ {\rm GeV} $ (heavy, fast).

Table \ref{tab:tildem_m0} shows the ratio of the effective mass and common DM masses, $\tilde{m}_{\rm DM}/m_0$ for $N_{\rm DM}=3, 5, 10, 100$ and $s=0, 1/2, 1, 3/2, 2$. Note that the analytic formulas, such as Eq. (\ref{Eq:Y_chi_slow}), provide good solutions for $\sum g_i \ll g_{\rm DM}$ \cite{Lennon2018JCAP}. Therefore, the predictions for  $N_{\rm DM} = 100$ may need a correction. Table \ref{tab:dm_mass_degenerated} shows the allowed region of DM mass for the degenerated DM mass spectrum for $N_{\rm DM}=3$ and $10$. The allowed regions of DM mass for $N_{\rm DM}=10$ are wider than those for $N_{\rm DM}=3$. For the light DM case, the upper limit of DM mass  increases with  $N_{\rm DM}$. In contrast, the lower limit of DM mass decreases with $N_{\rm DM}$ for the heavy DM case. 

%=============================================
\begin{table}[t]
\caption{Ratio of the effective DM mass and common DM masses, $\tilde{m}_{\rm DM}/m_0$.}
\small
\begin{center}
\begin{tabular}{|c|c|cccc|}
\hline
$s$& $N_{\rm DM}$  & Light and slow & Light and fast & Heavy and slow & Heavy and fast\\
\hline
0 & $3$  & $0.376  $ & $0.0472$ & $2.66 $ & $21.2$ \\
& $5$  & $0.618$ & $0.0763$ & $1.62$ & $13.1$ \\
& $10$  & $1.19$ & $0.142$ & $0.840$ & $7.05$ \\
& $100$  & $7.89$ & $0.623$ & $0.127$ & $1.60$ \\
\hline
$1/2$ & $3$  & $0.400$ & $0.0531$ & $2.51$ & $18.8$ \\
& $5$  & $0.653$ & $0.0854$ & $1.53$ & $11.7$ \\
& $10$  & $1.25$ & $0.157$ & $0.797$ & $6.35$ \\
& $100$  & $8.07$ & $0.651$ & $0.124$ & $1.54$ \\
\hline
$1$ & $3$  & $0.314$ & $0.0328$ & $3.19$ & $30.5$ \\
& $5$  & $0.517$ & $0.0535$ & $1.93$ & $18.7$ \\
& $10$  & $1.01$ & $0.102$ & $0.992$ & $9.85$ \\
& $100$  & $7.28$ & $0.531$ & $0.137$ & $1.88$ \\
\hline
$3/2$ & $3$  & $0.211$ & $0.0148$ & $4.74$ & $67.4$ \\
& $5$  & $0.350$ & $0.0250$ & $2.86$ & $40.8$ \\
& $10$  & $0.691$ & $0.0478$ & $1.45$ & $20.9$ \\
& $100$  & $5.78$ & $0.334$ & $0.173$ & $3.00$ \\
\hline
$2$ & $3$  & $0.140$ & $0.00653$ & $7.14$ & $153$ \\
& $5$  & $0.233$ & $0.0108$ & $4.30$ & $92.3$ \\
& $10$  & $0.463$ & $0.0214$ & $2.16$ & $46.6$ \\
& $100$  & $4.24$ & $0.180$ & $0.236$ & $5.56$ \\
\hline
\end{tabular}
\end{center}
\label{tab:tildem_m0}
\end{table}
%=============================================

%=============================================
\begin{table}[t]
\caption{Allowed region of DM mass [{\rm GeV}]  for the degenerated DM mass spectrum.}
\small
\begin{center}
\begin{tabular}{|c|c|cc|}
\hline
$s$& $N_{\rm DM}$  & Light and slow & Light and fast \\
\hline
0 & $3$  & $[2.6\times 10^{-7}, 0.30]$ & $[3.1\times 10^{-7}, 1.3\times 10^{12}]$ \\
& $10$  & $[3.1\times 10^{-7}, 0.80]$ & $[3.1\times 10^{-7}, 4.0\times 10^{12}]$ \\
\hline
$1/2$ & $3$  & $[3.6\times 10^{-7}, 0.44]$ & $[4.2\times 10^{-7}, 2.1\times 10^{12}]$ \\
& $10$  & $[4.5\times 10^{-7}, 1.1]$ & $[4.2\times 10^{-7}, 6.1\times 10^{12}]$ \\
\hline
$1$ & $3$  & $[7.8\times 10^{-7}, 0.75]$ & $[9.2\times 10^{-7}, 2.8\times 10^{12}]$ \\
& $10$  & $[7.9\times 10^{-7}, 2.4]$ & $[9.2\times 10^{-7}, 8.7\times 10^{12}]$ \\
\hline
$3/2$ & $3$  & $[2.0\times 10^{-6}, 1.3]$ & $[2.0\times 10^{-6}, 3.0\times 10^{12}]$ \\
& $10$  & $[2.0\times 10^{-6}, 4.1]$ & $[2.0\times 10^{-6}, 9.5\times 10^{12}]$ \\
\hline
$2$ & $3$  & $[6.3\times 10^{-6}, 2.7]$ & $[7.4\times 10^{-6}, 4.4\times 10^{12}]$\\
& $10$  & $[6.3\times 10^{-6}, 8.8]$ & $[7.4\times 10^{-6}, 1.5\times 10^{13}]$ \\
\hline
\hline
$s$& $N_{\rm DM}$  & Heavy and slow & Heavy and fast\\
\hline
0 & $3$  & $[9.0\times 10^{9}, 1.2\times 10^{19}]$ & $[6.1\times 10^{10}, 1.2\times 10^{19}]$ \\
& $10$  & $[3.4\times 10^{9}, 1.0\times 10^{19}]$ & $[2.0\times 10^{10}, 1.2\times 10^{19}]$ \\
\hline
$1/2$ & $3$  & $[7.8\times 10^{9}, 1.2\times 10^{19}]$ & $[4.9\times 10^{10}, 1.2\times 10^{19}]$ \\
& $10$  & $[3.1\times 10^{9}, 9.7\times 10^{18}]$ & $[1.7\times 10^{10}, 1.2\times 10^{19}]$ \\
\hline
$1$ & $3$ & $[3.5\times 10^{9}, 1.2\times 10^{19}]$ & $[2.9\times 10^{10}, 1.2\times 10^{19}]$ \\
& $10$  & $[1.1\times 10^{9}, 1.2\times 10^{19}]$ & $[9.5\times 10^{9}, 1.2\times 10^{19}]$ \\
\hline
$3/2$ & $3$  & $[2.4\times 10^{9}, 1.2\times 10^{19}]$ & $[3.4\times 10^{10}, 1.2\times 10^{19}]$ \\
& $10$  & $[7.3\times 10^{8}, 1.2\times 10^{19}]$ & $[1.0\times 10^{10}, 1.2\times 10^{19}]$ \\
\hline
$2$ & $3$  & $[1.0\times 10^{9}, 1.2\times 10^{19}]$ & $[1.8\times 10^{10}, 1.2\times 10^{19}]$\\
& $10$ & $[3.0\times 10^{8}, 1.2\times 10^{19}]$ & $[5.6\times 10^{9}, 1.2\times 10^{19}]$ \\
\hline
\end{tabular}
\end{center}
\label{tab:dm_mass_degenerated}
\end{table}
%=============================================

%%----------------------------------------------------------------------------------
\subsection{Arithmetic sequence-like DM mass spectrum}
%%----------------------------------------------------------------------------------
For the arithmetic sequence-like DM mass spectrum, for example, we obtain
\begin{eqnarray}
 \tilde{m}_{\rm DM} =\left( \frac{e_{\rm T, \chi}}{e_{\rm T}} \right)^{1/2} \left\{ m_0  \sum_{i=1}^{N_{\rm DM}} \left(1- \left(\frac{i-1}{N_{\rm DM}-1}\right)^p\right) + M \sum_{i=1}^{N_{\rm DM}} \left(\frac{i-1}{N_{\rm DM}-1}\right)^p\right\},
\end{eqnarray}
for the light and slow case. For $N_{\rm DM}=3$ and $s=0$, the effective DM mass is related to the minimum and maximum masses, $m_0$ and $M$, in the arithmetic sequence-like DM mass spectrum as follows
\begin{eqnarray}
\tilde{m}_{\rm DM}=
\begin{cases}
0.145 m_0 + 0.231M  & (p=1/4), \\
0.188m_0 + 0.188 M   & (p=1), \\
0.243m_0 + 0.133M  & (p=4). \\
\end{cases}
\end{eqnarray}

Similar to the case of degenerated DM mass spectrum case, we require that the relations
\begin{eqnarray}
m_\chi^{\rm min} <  m_0 < M < m_\chi^{\rm max}, \quad
m_\chi^{\rm min} <  \tilde{m}_{\rm DM} < m_\chi^{\rm max},
\end{eqnarray}
should be satisfied for the arithmetic sequence-like DM mass spectrum. For example, the allowed region of DM mass for $N_{\rm DM}=3$ and $s=0$ is given by
\begin{eqnarray}
\tilde{m}_{\rm DM}[{\rm GeV}] &\simeq&
\begin{cases}
2.616\times 10^{-7}  - 0.2929  & (p=1/4), \\
2.600\times 10^{-7}  - 0.2912  & (p=1), \\
2.603\times 10^{-7}  - 0.2891 & (p=4). \\
\end{cases}
\end{eqnarray}
If $\tilde{m}_{\rm DM}$ is calculated with two significant digits similar to Tab.\ref{tab:limit_mchi}, we obtain
\begin{eqnarray}
\tilde{m}_{\rm DM}[{\rm GeV}] &\simeq& 2.6\times 10^{-7}  - 0.29,
\end{eqnarray}
for $p=1/4$, $p=1$ and $p=4$, and there is no significant p-dependence. 

Tables \ref{tab:A_N3} and \ref{tab:A_N10} show the allowed region of DM mass for the arithmetic DM mass spectrum when  $N_{\rm DM}=3$ and $N_{\rm DM}=10$, respectively. Similar to the degenerated DM mass spectrum case, the allowed regions of DM mass for $N_{\rm DM}=10$ are wider than those for $N_{\rm DM}=3$. For the light DM case, the upper limit of DM mass increases with $N_{\rm DM}$. In contrast, the lower limit of DM mass decreases with  $N_{\rm DM}$ for the heavy DM case. 

%
%=============================================
\begin{table}[h]
\caption{Allowed region of DM mass [{\rm GeV}]  for the arithmetic DM mass spectrum for $N_{\rm DM}=3$.}
\small
\begin{center}
\begin{tabular}{|c|c|cc|}
\hline
 $s$ & $p$ & Light and slow & Light and fast \\ 
\hline
 & $1/4$ & $[2.6 \times 10^{-7}, 0.29]$  & $[3.1 \times 10^{-7}, 1.3 \times 10^{12}]$\\
0 & $1$ & $[2.6 \times 10^{-7}, 0.29]$  & $[3.1 \times 10^{-7}, 1.2 \times 10^{12}]$\\
 & $4$ & $[2.6 \times 10^{-7}, 0.29]$  & $[3.1 \times 10^{-7}, 1.2 \times 10^{12}]$\\
\hline
 & $1/4$ & $[3.6 \times 10^{-7}, 0.43]$  & $[4.2 \times 10^{-7}, 2.0 \times 10^{12}]$\\
1/2 & $1$ & $[3.6 \times 10^{-7}, 0.43]$  & $[4.2 \times 10^{-7}, 2.0 \times 10^{12}]$\\
 & $4$ & $[3.6 \times 10^{-7}, 0.42]$  & $[4.2 \times 10^{-7}, 2.0 \times 10^{12}]$\\
\hline
 & $1/4$ & $[7.8 \times 10^{-7}, 0.73]$  & $[9.2 \times 10^{-7}, 2.7 \times 10^{12}]$\\
1 & $1$ & $[7.8 \times 10^{-7}, 0.73]$  & $[9.2 \times 10^{-7}, 2.7 \times 10^{12}]$\\
 & $4$ & $[7.8 \times 10^{-7}, 0.72]$  & $[9.2 \times 10^{-7}, 2.7 \times 10^{12}]$\\
\hline
 & $1/4$ & $[2.0 \times 10^{-6}, 1.2]$  & $[2.0 \times 10^{-6}, 2.8 \times 10^{12}]$\\
3/2 & $1$ & $[2.0 \times 10^{-6}, 1.2]$  & $[2.0 \times 10^{-6}, 2.8 \times 10^{12}]$\\
 & $4$ & $[2.0 \times 10^{-6}, 1.2]$  & $[2.0 \times 10^{-6}, 2.8 \times 10^{12}]$\\
\hline
 & $1/4$ & $[6.3 \times 10^{-6}, 2.5]$  & $[7.4 \times 10^{-6}, 4.3 \times 10^{12}]$\\
2 & $1$ & $[6.3 \times 10^{-6}, 2.5]$  & $[7.4 \times 10^{-6}, 4.3 \times 10^{12}]$\\
 & $4$ & $[6.3 \times 10^{-6}, 2.5]$  & $[7.4 \times 10^{-6}, 4.3 \times 10^{12}]$\\
\hline
\hline
 $s$ & $p$ & Heavy and slow & Heavy and fast \\ 
\hline
 & $1/4$ & $[9.8 \times 10^{9}, 1.2 \times 10^{19}]$  & $[6.6 \times 10^{10}, 1.2 \times 10^{19}]$\\
0 & $1$ & $[9.7 \times 10^{9}, 1.2 \times 10^{19}]$  & $[6.6 \times 10^{10}, 1.2 \times 10^{19}]$\\
 & $4$ & $[9.7 \times 10^{9}, 1.2 \times 10^{19}]$  & $[6.6 \times 10^{10}, 1.2 \times 10^{19}]$\\
\hline
 & $1/4$ & $[8.4 \times 10^{9}, 1.2 \times 10^{19}]$  & $[5.3 \times 10^{10}, 1.2 \times 10^{19}]$\\
1/2 & $1$ & $[8.4 \times 10^{9}, 1.2 \times 10^{19}]$  & $[5.3 \times 10^{10}, 1.2 \times 10^{19}]$\\
 & $4$ & $[8.3 \times 10^{9}, 1.2 \times 10^{19}]$  & $[5.2 \times 10^{10}, 1.2 \times 10^{19}]$\\
\hline
 & $1/4$ & $[3.8 \times 10^{9}, 1.2 \times 10^{19}]$  & $[3.2 \times 10^{10}, 1.2 \times 10^{19}]$\\
1 & $1$ & $[3.8 \times 10^{9}, 1.2 \times 10^{19}]$  & $[3.2 \times 10^{10}, 1.2 \times 10^{19}]$\\
 & $4$ & $[3.8 \times 10^{9}, 1.2 \times 10^{19}]$  & $[3.1 \times 10^{10}, 1.2 \times 10^{19}]$\\
\hline
 & $1/4$ & $[2.6 \times 10^{9}, 1.2 \times 10^{19}]$  & $[3.6 \times 10^{10}, 1.2 \times 10^{19}]$\\
3/2 & $1$ & $[2.6 \times 10^{9}, 1.2 \times 10^{19}]$  & $[3.6 \times 10^{10}, 1.2 \times 10^{19}]$\\
 & $4$ & $[2.5 \times 10^{9}, 1.2 \times 10^{19}]$  & $[3.6 \times 10^{10}, 1.2 \times 10^{19}]$\\
\hline
 & $1/4$ & $[1.1 \times 10^{9}, 1.2 \times 10^{19}]$  & $[2.0 \times 10^{10}, 1.2 \times 10^{19}]$\\
2 & $1$ & $[1.1 \times 10^{9}, 1.2 \times 10^{19}]$  & $[2.0 \times 10^{10}, 1.2 \times 10^{19}]$\\
 & $4$ & $[1.1 \times 10^{9}, 1.2 \times 10^{19}]$  & $[2.0 \times 10^{10}, 1.2 \times 10^{19}]$\\
\hline
\end{tabular}
\end{center}
\label{tab:A_N3}
\end{table}
%=============================================

%
%=============================================
\begin{table}[t]
\caption{Allowed region of DM mass [{\rm GeV}] for the arithmetic DM mass spectrum for $N_{\rm DM}=10$.}
\small
\begin{center}
\begin{tabular}{|c|c|cc|}
\hline
 $s$ & $p$ & Light and slow & Light and fast \\ 
\hline
 & $1/4$ & $[3.4 \times 10^{-7}, 0.8]$  & $[3.1 \times 10^{-7}, 3.8 \times 10^{12}]$\\
0 & $1$ & $[3.3 \times 10^{-7}, 0.8]$  & $[3.1 \times 10^{-7}, 3.7 \times 10^{12}]$\\
 & $4$ & $[3.3 \times 10^{-7}, 0.8]$  & $[3.1 \times 10^{-7}, 3.7 \times 10^{12}]$\\
\hline
 & $1/4$ & $[4.9 \times 10^{-7}, 1.1]$  & $[4.2 \times 10^{-7}, 6.0 \times 10^{12}]$\\
1/2 & $1$ & $[4.9 \times 10^{-7}, 1.1]$  & $[4.2 \times 10^{-7}, 5.9 \times 10^{12}]$\\
 & $4$ & $[4.8 \times 10^{-7}, 1.1]$  & $[4.2 \times 10^{-7}, 5.8 \times 10^{12}]$\\
\hline
 & $1/4$ & $[8.6 \times 10^{-7}, 2.4]$  & $[9.2 \times 10^{-7}, 8.4 \times 10^{12}]$\\
1 & $1$ & $[8.5 \times 10^{-7}, 2.3]$  & $[9.2 \times 10^{-7}, 8.3 \times 10^{12}]$\\
 & $4$ & $[8.4 \times 10^{-7}, 2.3]$  & $[9.2 \times 10^{-7}, 8.2 \times 10^{12}]$\\
\hline
 & $1/4$ & $[2.0 \times 10^{-6}, 4.0]$  & $[2.0 \times 10^{-6}, 9.1 \times 10^{12}]$\\
3/2 & $1$ & $[2.0 \times 10^{-6}, 3.9]$  & $[2.0 \times 10^{-6}, 9.0 \times 10^{12}]$\\
 & $4$ & $[2.0 \times 10^{-6}, 3.9]$  & $[2.0 \times 10^{-6}, 8.8 \times 10^{12}]$\\
\hline
 & $1/4$ & $[6.3 \times 10^{-6}, 8.4]$  & $[7.4 \times 10^{-6}, 1.4 \times 10^{13}]$\\
2 & $1$ & $[6.3 \times 10^{-6}, 8.3]$  & $[7.4 \times 10^{-6}, 1.4 \times 10^{13}]$\\
 & $4$ & $[6.3 \times 10^{-6}, 8.2]$  & $[7.4 \times 10^{-6}, 1.4 \times 10^{13}]$\\
\hline
\hline
 $s$ & $p$ & Heavy and slow & Heavy and fast \\ 
\hline
 & $1/4$ & $[3.4 \times 10^{9}, 9.7 \times 10^{18}]$  & $[2.2 \times 10^{10}, 1.2 \times 10^{19}]$\\
0 & $1$ & $[3.5 \times 10^{9}, 9.6 \times 10^{18}]$  & $[2.2 \times 10^{10}, 1.2 \times 10^{19}]$\\
 & $4$ & $[3.5 \times 10^{9}, 9.4 \times 10^{18}]$  & $[2.2 \times 10^{10}, 1.2 \times 10^{19}]$\\
\hline
 & $1/4$ & $[3.1 \times 10^{9}, 9.2 \times 10^{18}]$  & $[1.8 \times 10^{10}, 1.2 \times 10^{19}]$\\
1/2 & $1$ & $[3.2 \times 10^{9}, 9.1 \times 10^{18}]$  & $[1.8 \times 10^{10}, 1.2 \times 10^{19}]$\\
 & $4$ & $[3.1 \times 10^{9}, 9.0 \times 10^{18}]$  & $[1.8 \times 10^{10}, 1.2 \times 10^{19}]$\\
\hline
 & $1/4$ & $[1.2 \times 10^{9}, 1.2 \times 10^{19}]$  & $[1.0 \times 10^{10}, 1.2 \times 10^{19}]$\\
1 & $1$ & $[1.2 \times 10^{9}, 1.2 \times 10^{19}]$  & $[1.0 \times 10^{10}, 1.2 \times 10^{19}]$\\
 & $4$ & $[1.2 \times 10^{9}, 1.2 \times 10^{19}]$  & $[1.0 \times 10^{10}, 1.2 \times 10^{19}]$\\
\hline
 & $1/4$ & $[7.9 \times 10^{8}, 1.2 \times 10^{19}]$  & $[1.1 \times 10^{10}, 1.2 \times 10^{19}]$\\
3/2 & $1$ & $[7.8 \times 10^{8}, 1.2 \times 10^{19}]$  & $[1.1 \times 10^{10}, 1.2 \times 10^{19}]$\\
 & $4$ & $[7.7 \times 10^{8}, 1.2 \times 10^{19}]$  & $[1.1 \times 10^{10}, 1.2 \times 10^{19}]$\\
\hline
 & $1/4$ & $[3.3 \times 10^{8}, 1.2 \times 10^{19}]$  & $[6.1 \times 10^{9}, 1.2 \times 10^{19}]$\\
2 & $1$ & $[3.3 \times 10^{8}, 1.2 \times 10^{19}]$  & $[6.0 \times 10^{9}, 1.2 \times 10^{19}]$\\
 & $4$ & $[3.2 \times 10^{8}, 1.2 \times 10^{19}]$  & $[5.9 \times 10^{9}, 1.2 \times 10^{19}]$\\
\hline
\end{tabular}
\end{center}
\label{tab:A_N10}
\end{table}
%=============================================

%%----------------------------------------------------------------------------------
\subsection{Geometric sequence-like DM mass spectrum}
%%----------------------------------------------------------------------------------
Because of the similarity between the arithmetic sequence-like DM mass spectrum and geometric sequence-like DM mass spectrum cases as addressed in Section  \ref{section:PBHandDM}, almost same allowed region of DM mass was obtained in the geometric and arithmetic DM mass spectrum cases.

For example,  the allowed region of DM mass for $N_{\rm DM}=3$ and $s=0$ is obtained as
\begin{eqnarray}
\tilde{m}_{\rm DM}[{\rm GeV}] &\simeq&
\begin{cases}
2.607\times 10^{-7}  - 0.2928  & (p=1/4), \\
2.601\times 10^{-7}  - 0.2911  & (p=1), \\
2.600\times 10^{-7}  - 0.2891 & (p=4), \\
\end{cases}
\end{eqnarray}
for the geometric sequence-like DM mass spectrum case. If $\tilde{m}_{\rm DM}$ is calculated with two significant digits similar to Table \ref{tab:limit_mchi}, the same results shown in Table \ref{tab:A_N3} are obtained even if in the case of a geometric sequence-like DM mass spectrum. Thus, showing all the results for the geometric sequence-like DM mass spectrum case was avoided in this paper.

%%----------------------------------------------------------------------------------
\section{Summary\label{sec:summary}}
%%----------------------------------------------------------------------------------
In principle, several DM species with specific mass spectra may be emitted from PBHs because PBHs evaporate into all particle species in nature. In addition, PBHs were assumed to be the only source of DMs, and DMs only interact with the standard model particles gravitationally. The effects of DM mass spectra on PBHs and vice versa have been investigated. The following four specific mass spectra of the DM sector have been considered: (a) single DM with a single mass; (b) multiple DMs with a degenerated mass spectrum; (c) multiple DMs with an arithmetic sequence-like mass spectrum; (d) multiple DMs with a geometric sequence-like mass spectrum.

The first finding shows that $\beta$-$M_{\rm BH}^{\rm in}$ curves for different numbers of DM species $N_{\rm DM}$ are distinguishable for degenerated DM mass spectra. On the other hand, for multiple DMs with an arithmetic sequence-like or a geometric sequence-like mass spectrum, $\beta$-$M_{\rm BH}^{\rm in}$ curves for different $N_{\rm DM}$ tends to overlap with each other for heavy initial PBHs. It can be concluded that $\beta$-$M_{\rm BH}^{\rm in}$ curves for different $N_{\rm DM}$ may tend to overlap with each other for heavy initial PBHs regardless of the derailed structure because of the similarity between the arithmetic and geometric sequence-like DM mass spectra. Currently, both the DM spectra and PBH parameters are independent and free parameters. If the initial density $\beta$ and initial mass $M_{\rm BH}^{\rm in}$ of PBHs are determined precisely in future experiments and reliable connections between PBHs and multiple DMs are obtained in future theoretical studies, $N_{\rm DM}$, $\beta$ and $M_{\rm BH}^{\rm in}$ may be connected. Even if that is the case, the determination of $N_{\rm DM}$ using $\beta$ and $M_{\rm BH}^{\rm in}$ may be difficult for heavy initial PBHs because of the existence of the overlap of $\beta$-$M_{\rm BH}^{\rm in}$ curves. 
 
Furthermore, according to Lennon, et al., \cite{Lennon2018JCAP}, multiple DMs were classified into four categories: (1) light and slow, (2) light and fast, (3) heavy and slow, and (4) heavy and fast. The good analytic solutions of the DM mass were used for these four categories. Consequently, the DM masses for multiple DM were determined to be more constrained than it in the single DM case. For the light DM case, the upper limit of the DM masses increases with $N_{\rm DM}$. On the other hand, the lower limit of the DM mass decreases with $N_{\rm DM}$, for the heavy DM case.

%\section*{Acknowledgment}

%Insert the Acknowledgment text here.


%\bibliographystyle{ptephy}
%\bibliography{sample}
%
% once the .bbl file has been generated then place the text in your article.

\vspace{0.2cm}
\noindent
%For references,  note how to include DOI information from examples below. 

%This is added by T. Yoneya (editor-in-chief) on 2020/07/09.

%\let\doi\relax

%without this code before the command "\begin{thebibliography}{}" , an error will be %flagged. When the bibliography is provided as separate .bib file, then this code %should be placed above the commands "\bibliographystyle{}" and "\bibliography{}" %inside the main TeX file. 

\end{document}